\documentclass[11pt]{article}
\usepackage[utf8]{inputenc}
\usepackage{amsmath,setspace,geometry}
\usepackage{amsfonts}
\usepackage[shortlabels]{enumitem}
\usepackage{graphicx}
\usepackage{bbm}

\usepackage[colorlinks,citecolor=purple,urlcolor=blue,bookmarks=false,hypertexnames=true]{hyperref}
\usepackage[]{natbib} 
\bibpunct[:]{(}{)}{,}{a}{}{,}
\usepackage[english]{babel}
\usepackage{float}
\usepackage{booktabs}
\usepackage{pdfpages}
\usepackage{threeparttable}
\usepackage{lscape}
\usepackage{subfig}

\title{Unified Container Shipping Industry Data From 1966: Freight Rate, Shipping Quantity, Newbuilding, Secondhand, and Scrap Price\thanks{ Declarations of interest: none}}
\author{Takuma Matsuda\thanks{Faculty of Commerce, Takushoku University. Email: tmatsuda@ner.takushoku-u.ac.jp}\quad Suguru Otani\thanks{Department of Economics, Rice University. Email: so19@rice.edu}}
\date{
First version: November 29, 2022\\
Current version: \today
}

\begin{document}

\maketitle

\begin{abstract}
We construct a new unified panel dataset that combines route-year-level freight rates with shipping quantities for the six major routes and industry-year-level newbuilding, secondhand, and scrap prices from 1966 (the beginning of the industry) to 2009. We offer detailed instructions on how to merge various datasets and validate the data's consistency by industry experts and former executives who have historical knowledge and experience. Using this dataset, we provide a quantitative and descriptive analysis of the industry dynamics known as the container crisis. Finally, we identify structural breaks for each variable to demonstrate the impact of the shipping cartels' collapse.
\end{abstract} 

\vspace{0.1in}
\noindent\textbf{Keywords:} container shipping industry; exemption agreement; container crisis; shipping cartel; container freight rate 
\vspace{0in}


{\flushright{``When we think about technology that changes the world, we think about glamorous things like the Internet. But if you try to figure out what happened to world trade, there is a strong case to be made that it was the container ...”\\
-- Krugman, Paul.\footnote{Citigroup Foundation Special Lecture, Festschrift paper in honor of Alan V. Deardorff, University of Michigan IPC working paper 91, 2009}\\
``[W]ithout [the container], the tremendous expansion of world trade in the last forty years—the fastest growth in any major economic activity ever recorded—could not possibly have taken place.”\\
-- Drucker, Peter F.\footnote{Innovation and Entrepreneurship, Butterworth-Heinemann, 2007 (p.28)}}
\flushright{}}

\section{Introduction}\label{sec:introduction}
\subsection{Research background, motivation and purpose}
Container shipping is a crucial component of global trade that has revolutionized the world. According to IHS Markit and Descartes Datamyne, container shipping accounts for 45.4\% of amount-based imports to the U.S., 21.3\% of amount-based exports from the U.S., and 10.12\% of quantity-based world trade are shipped in 2021. Additionally, the container shipping industry offers a fascinating opportunity to explore industry dynamics. Since the industry began global shipping operations in 1966, the market's initial state can be determined with a significant entry and exit of firms. Furthermore, many countries have competition laws that exempt international cartels and consortia, which are called ``shipping conferences" and ``shipping alliances"\footnote{\cite{Anteitekina_kokusaikaijouyusoukakuhonotameno_kaijiseisakuno_arikatanitsuite2007} of the Ministry of Land, Infrastructure and Transportation in Japan defines a shipping conference as an agreement on freight rates and other business matters to restrain competition among container shipping lines along the same route. A consortium is a technical agreement between several shipping lines to form a group and operate a joint service for the diversification of services and cost reduction in the liner service. Consortiums include alliances that are widely used. } in the shipping industry.\footnote{There is an important difference between shipping conferences and alliances: conferences are organized by specific routes and directions, while alliances are formed globally. There may exist one conference covering trade on the Transatlantic route and another covering trade between Northern Europe and the U.S. Gulf ports. In addition, firms do not always participate in conferences on all routes and directions they serve. Thus, conferences are heterogeneous in their structure and membership.} Despite its significance, there is a lack of a panel dataset for the route-year-level freight rate and shipping quantity in the container shipping industry, particularly for the years 1966 to 1990, which limits quantitative research during this period\footnote{As other related studies focus on the container freight rate panel data, \cite{luo2009econometric} use shipping demand and freight rate data between 1980 and 2007. As shipping demand, they used the world container throughput reported in the Drewry Annual Container Market Review and Forecast. The container freight rate is calculated as the weighted average of the Transpacific, Europe-Far East, and Transatlantic trades from the same data source. Due to data limitations, they calculated the missing period (1980–1993) from the General Freight Index in the Shipping Statistics Yearbook 2007, using a simple statistical equation between the container freight rate and the general freight index from 1994 to 2008.}. This study provides a unified dataset based on published books and publicly available data sources to the extent possible, combining route-year-level data with new industry-level data regarding shipbuilding, secondhand, and scrap prices.\footnote{The data sources are listed in Appendix \ref{sec:data_construction}.}

To construct the freight index, we adopted the tractable imputation approach, which involves linking multiple data sources that overlap information for some years. This approach is simpler than formal but complicated processes or estimation approaches that assume the AR(1) process under stationary assumptions \citep{jeon2022learning}. We also validated our approach by presenting interview-based evidence from ex-executive officers of shipping companies. This confirms that our Twenty-foot Equivalent Unit (TEU)-based price data provide a reasonable benchmark measure, even though container freight rates were not determined by container units in the 1970s.

Using our new dataset, we have conducted an analysis of the historical shipping price reductions in the 1980s, known as the ``container crisis" \citep{broeze2002globalisation}, by implementing the unknown multiple structural breaks test \citep{bai1998estimating,bai2003computation}. It is anecdotally known that the crisis was triggered by two events: (1) the withdrawal of Sea-Land, which was the biggest cartel member from shipping cartels in 1980, and (2) the enactment of the Shipping Act of 1984. The first event changed the market share of shipping cartels, whereas the second event neutralized the shipping cartels. We also applied the test to industry-year-level newbuilding, secondhand, and scrap prices to determine the relationship between shipping markets and shipbuilding markets.

Based on the test, we found that the container crisis occurred on all six routes by 1980, which was triggered by the withdrawal of Sea-Land, although non-U.S. routes reacted earlier than U.S. routes. Additionally, we found that industry-year-level newbuilding, secondhand, and scrap prices displayed a relatively stable trend during the container crisis, unlike shipping prices. We interpret these differences as naturally capturing the distinctions between shipping markets with shipping conferences and shipbuilding markets without cartel groups. Thus, the container crisis was a specific event in the shipping market.

\subsection{Literature review}\label{subsec:litereture}

This study contributes to two strands of the literature: the historical connection to the recent growing empirical research on the shipping industry and the effect of an explicit cartel on the price in the shipping industry.

First, this study provides the necessary data to connect the history of the container shipping industry from its beginning to its development after 2000, a topic that has gained attention in the industrial organization literature \citep{aguirregabiria2021dynamic}. The most relevant paper in this regard is \cite{jeon2022learning}, in which she examined the relationship between demand uncertainty and firm-market-year-level investment decisions using container shipping demand and freight rate data for the years 1997 to 2014. To facilitate her learning-based model, she needed to obtain the initial prices and demand in the container industry. However, due to data limitations, she had to resort to imputation and truncation approaches for missing shipping demand data from 1966: Q2 to 1996: Q4.\footnote{On cartel issues in a similar industry, the most relevant paper is by \cite{asker2010leniency}. He studied how the presence of a cartel affects market conduct following its dissolution and how the dissolution might be affected by the obligations imposed on firms that seek leniency in the tanker shipping market between 2001 and 2002. \cite{kalouptsidi2014aer}, \cite{brancaccio2020geography}, and \cite{greenwood2015waves} investigated the bulk shipping industry after 2000. \cite{bai2021congestion} studied the tanker market during 2017-2020 and investigated how the imbalance between the demand for and the supply of shipping services
determines congestion. Although these industries are closely related to the container shipping industry, each of the industries is characterized differently because the market structure and competition are different. In addition, \cite{kalouptsidi2017res} and \cite{barwick2019china} studied the shipbuilding industry, which is an upstream industry for the container and bulk shipping industries. These papers rely on the use of Clarksons Research database and focus on the period after 2000. In the literature on international trade, \cite{bernhofen2016estimating} use country-level panel data regarding containerization adoption for the period 1962–1990. They focused on the effect of containerization adoption on the trade quantity; therefore, they refer to \textit{the Containerization International Yearbook} only for obtaining the first presence of container technology in each country. \cite{rua2014diffusion} added port-level data to their study and investigated the diffusion of initial adoptions of containerized transportation. \cite{cocsar2018shipping} exploited rich Turkish export data and examined modal choice between containerization and breakbulk shipping.} Our study overcomes this limitation and complements the findings of previous studies.

Second, this study examines the effect of explicit shipping cartels on shipping prices.\footnote{
The ongoing study of one of the authors of this paper investigates the effect by constructing a structural model and focuses on the shipping conferences as a single explicit cartel entity, and departs from the traditional focus on tacit collusion in the literature, e.g., \cite{porter1983study}, \cite{bresnahan1987competition}, \cite{miller2017understanding}, and \cite{byrne2019learning}.
The institutional background is similar to \cite{igami2015market}'s findings. He studied the impact of market power on international coffee prices and evaluated the impact of a cartel treaty on coffee prices and its global welfare consequences under counterfactual competition regimes, that is, collusion versus Cournot–Nash in a single homogenous good market.} Shipping cartels have been recognized in survey papers, such as \cite{levenstein2006determines} and \cite{asker2021}. Some studies, such as \cite{morton1997entry} and \cite{podolny1999social}, examine specific shipping cartels in the U.K. in the 1800s. The most relevant studies are \cite{wilson1991some}, \cite{pirrong1992application}, and \cite{clyde1998market}. \cite{wilson1991some} provided evidence of regime change by the Shipping Act of 1984 using data on quarterly freight rates and shipping quantities of five selected commodities only on the Transpacific route. \cite{pirrong1992application} tested the model prediction of the core theory surveyed in \cite{sjostrom2013competition} using data from two specific trade routes between 1983 and 1985. \cite{clyde1998market} studied the relationship between market power and the market share of shipping conferences after the act. However, they did not exploit cross-sectional variations, especially between non-U.S. and U.S. routes. Our study is the first empirical research to detect the effect of shipping cartels on the six main routes after global containerization in the 1970s, and it overcomes data limitations to complement the findings of previous studies.

The remainder of this paper is organized as follows. Section \ref{sec:data} summarizes the data and institutional background of the container shipping industry. Section \ref{sec:interview} presents interview-based evidence from an ex-chairperson and an ex-vice chairperson to demonstrate the consistency of our recovered data. Section \ref{sec:structural_break_test} implements structural break tests to detect price reductions in market variables. Finally, Section \ref{sec:conclusion} presents our conclusions, and Appendix \ref{sec:data_construction} presents the details of data construction.

\section{Data and Institutional Details}\label{sec:data}

We provide details of the data source in Section \ref{subsec:data}. Next, we provide graphical interpretations in Section \ref{subsec:graphical_interpretation} and summary statistics for the variables in Section \ref{subsec:summary_statistics}. We also introduce institutional details. For clarification, in this industry, a market refers to a non-directional location pair. For instance, if a firm operates a container ship in the eastbound of the transpacific, it should also operate the ship in the westbound. A \textit{route} refers to a directional round-trip between two locations. For example, both the westbound and eastbound of the transpacific. An \textit{industry} refers to the entire market consisting of all six routes.

\subsection{Data source}\label{subsec:data}
We constructed route-year-level and industry-year-level data for the container shipping industry.\footnote{Based on the firm-level data, each route is divided into conference and non-conference routes. For example, the Transatlantic eastbound conference route is a single route and the Transatlantic conference market is a single market. A conference market is a market where all conference firms conducted collusive behavior under the shipping conferences before the Shipping Act, but have competed since the act, whereas a non-conference market is a market where all non-conference firms have competed throughout the whole period. In the interview with ex-practitioners in shipping companies, they told that container freight rates for nonmember carriers are 20\% to 30\% lower than conference firms on the same routes. However, as far as we know, there was no data available on the trends in freight rates offered by the non-conference carriers. In addition, although we checked \textit{The Japan Maritime Daily}, the newspaper of the maritime industry, we could not find any articles that continuously reported on the level of freight rates offered by the non-conference carriers. } First, we collected route-year-level data on container freight rates and shipping quantities. Collecting data, particularly before 1994, was not trivial because no single data source covers the period from 1966 to 1993. Thus, we needed to carefully combine data from multiple sources carefully.\footnote{For reference, we merge \textit{``Issues of Our Ocean Shipping"} (\textit{``Wagakuni no Gaikou Kaiun Ni Tsuite," in Japanese}), \textit{Global Container Markets Drewry Shipping Consultants}, \textit{Review of Maritime Transport}, \textit{Containerization International 1973}, \textit{World Sea Trade Service}, \textit{Container transportation cost and profitability 1980/2000}, \textit{The Container Crisis 1982}, \textit{World Container Data 1985}, and \textit{World's sea trades}. Then, we converted merged data into TEU-based data.} In Appendix \ref{sec:data_construction}, we provide the data source and a detailed guide with some assumptions on data construction for each container freight rate and shipping quantity on the six major routes: mainhaul and backhaul (separately) on the Transpacific (Asia and North America), Transatlantic (North America and Europe), and Asia-Europe routes. Finally, we used price and quantity data from the six routes between 1966 and 2009. The price was adjusted according to the CPI in the U.S. in 1995. To check the accuracy and validity of our data, in Section \ref{sec:interview}, we provide interview-based evidence on the consistency of our recovered data with the historical experience of industry experts, Akimitsu Ashida and Hiroyuki Sato.

Second, we utilized industry-year-level data for newbuilding, secondhand, and scrap prices.\footnote{For reference, we merged \textit{Review (1971-1998)} and \textit{Lloyd's Shipping Economist (1983-1990)}. Then, we converted merged data into TEU-based data.} Specifically, we used newbuilding and scrap prices per TEU between 1966 and 1998 and the secondhand price per TEU between 1968 and 1998. These prices were adjusted to the CPI in the U.S. in 1995. 

It is worth noting that recent data on newbuilding, secondhand, and scrap prices after 1998 and shipping prices and quantities after 2009 are available for purchase from companies such as Clarksons and IHS Markit. While our data can be merged with proprietary data, this study does not include such data in order to provide public access to our data.

\subsection{Graphical interpretation}\label{subsec:graphical_interpretation}
\subsubsection{Route-year-level shipping price and quantity data}

\begin{figure}[!ht]
\begin{center}
  \subfloat[Price]{\includegraphics[width = 0.7\textwidth]
  {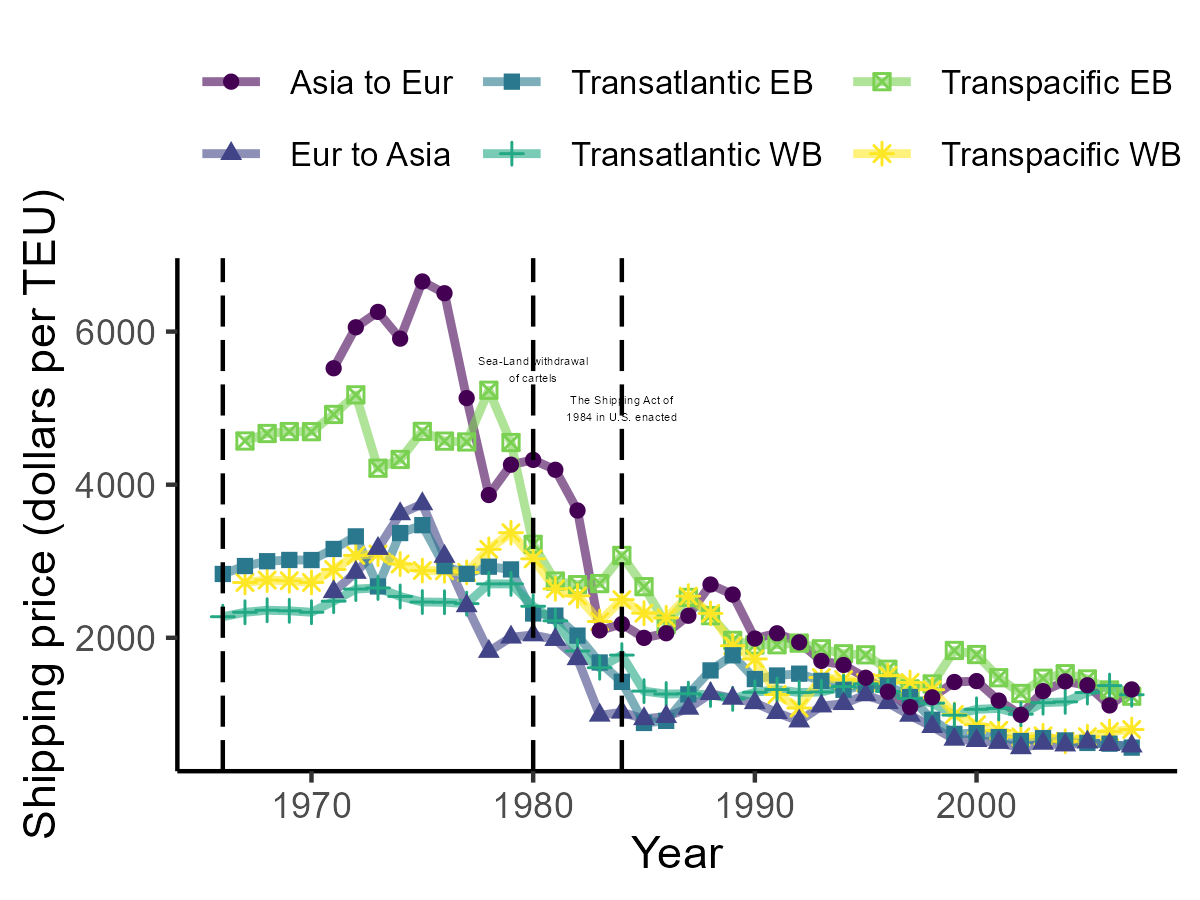}}\\
  \subfloat[Quantity]{\includegraphics[width = 0.7\textwidth]
  {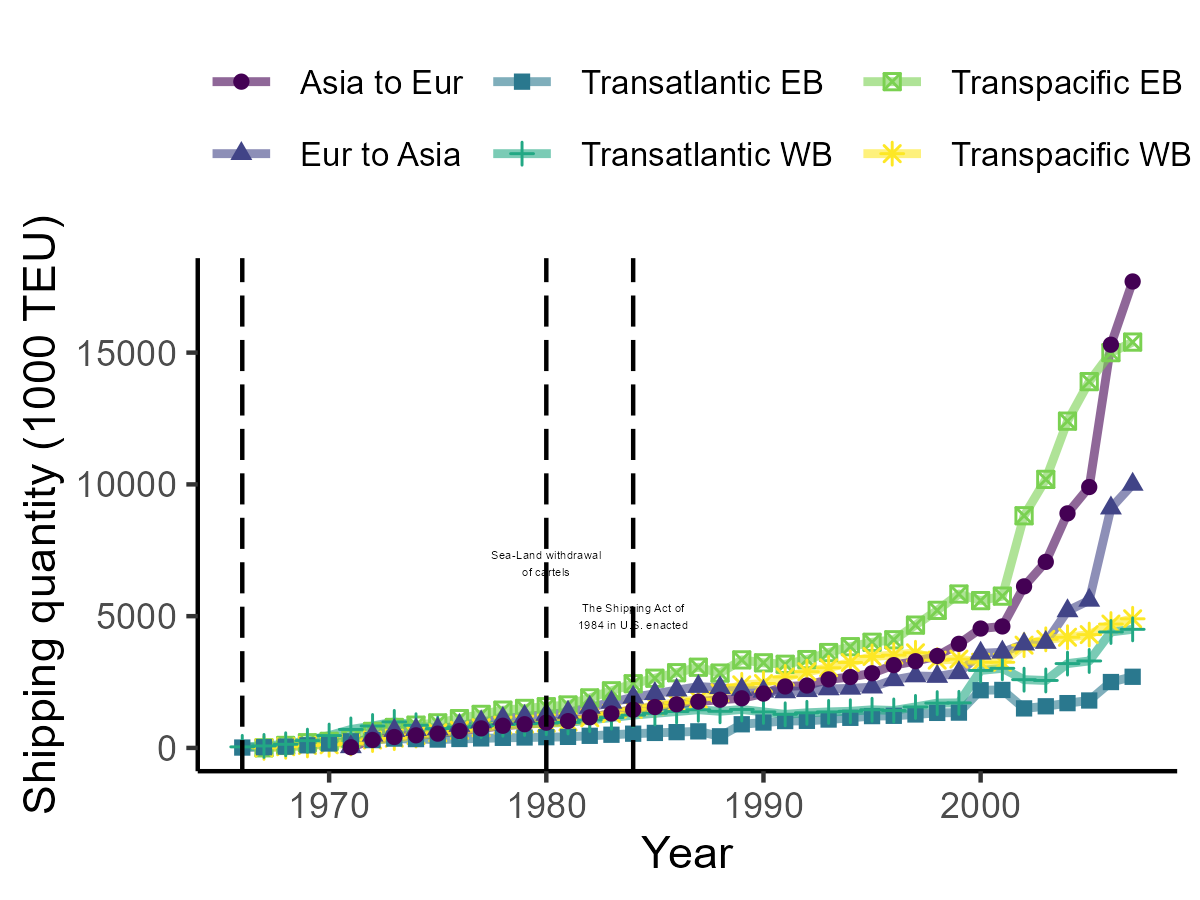}}
  \caption{Trends in route-year-level shipping prices and quantities.}
  \label{fg:container_freight_rate_and_shipping_quantity_each_route}
  \end{center}
\footnotesize
  Note: Container freight rates before 1992 refer to conference prices. The container freight rates after 1993 are unified prices based on conference and non-conference prices, and the difference between them is known to vanish because of the Shipping Act of 1984. The latter are standard data often used in the literature, such as \cite{jeon2022learning}. Prices were adjusted to the CPI in the U.S. in 1995.
\end{figure}

Figure \ref{fg:container_freight_rate_and_shipping_quantity_each_route} illustrates the nonstationary trends in container freight rates and shipping quantity between 1966 and 1990. The container freight rate decreased with fluctuations, and suddenly dropped significantly, with the transition of freight rates on the Asia-Europe eastbound and westbound routes being more unstable than those on Transpacific and Transatlantic routes in the 1970s and 1980s. This was mainly due to the reopening of the Suez Canal in 1976, which increased the supply of container shipping services \citep{jsme2022}. Additionally, freight rates for the Asia to Europe trade were higher than for other trade routes, possibly due to the strong influence of shipping conferences on Asia to Europe trade in the 1970s and early 1980s. The figure shows a sharp decline in freight rates in the second half of the 1980s, possibly because of the significant impact of the conferences' loss of power in the early 1980s.\footnote{In the interview with an ex-executive of a Japanese shipping company, he pointed out the difference between conferences. Conferences related to Asia-Europe had a strict membership screening process, and the number of voyages by member companies was clearly defined (closed conference). In addition, some members in the conference pooled their freight and then redistributed them. In contrast, the conferences for Transpacific routes were free to join or leave (open conference), and freight pooling was explicitly prohibited under the Shipping Act of 1916 in the U.S.}

The shipping quantity on all routes increased monotonically between 1973 and 1990 due to the increase in containerized cargo and the increase in the size of ships.\footnote{When Sea-land began container transport services in 1966, 226 35-foot containers were deployed. In the 1970s, with the development of international maritime container transport, container vessels began to increase in size, with vessels of approximately 2000 TEU becoming the mainstay of liner services. Owing to increased competition following the Shipping Act of 1984, shipping companies wanted to introduce even larger vessels. In 1988, American President Lines (APL) opened a 39.1-meter wide, 4300 TEU vessel, the first container vessel that could not pass the Panama Canal. Further, in the 1990s, 6000TEU-class vessels come into the container shipping market.} 
In response to the rapid increase in cargo exported from Asian countries, especially China, to Europe and the U.S., the number of vessels deployed on the Trans-Pacific East and Asia-Europe East routes has increased exponentially since 2000. \cite{jeon2022learning} has examined these characteristics in detail since 2000. 

The enactment of the Shipping Act of 1984 in the U.S. sharply divided the conference market regime into two periods, before and after 1984, as illustrated by the trend of the container freight rate in Figure \ref{fg:container_freight_rate_and_shipping_quantity_each_route}. The Shipping Act of 1984 in the U.S. made it much easier to form shipping conferences. As a result, after 1984, the Transpacific and Transatlantic routes saw significant price reductions due to market competition.\footnote{This is consistent with an interview article to ex-executives of Japanese shipping companies \citep{JapanMaritimeDaily2006} that were in charge of container trades in the 1980s.} The change in the market regime played a significant role in shaping the container crisis.

\subsubsection{Industry-year-level newbuilding, secondhand, and scrap prices data}

\begin{figure}[!ht]
\begin{center}
\includegraphics[width = 0.7\textwidth]{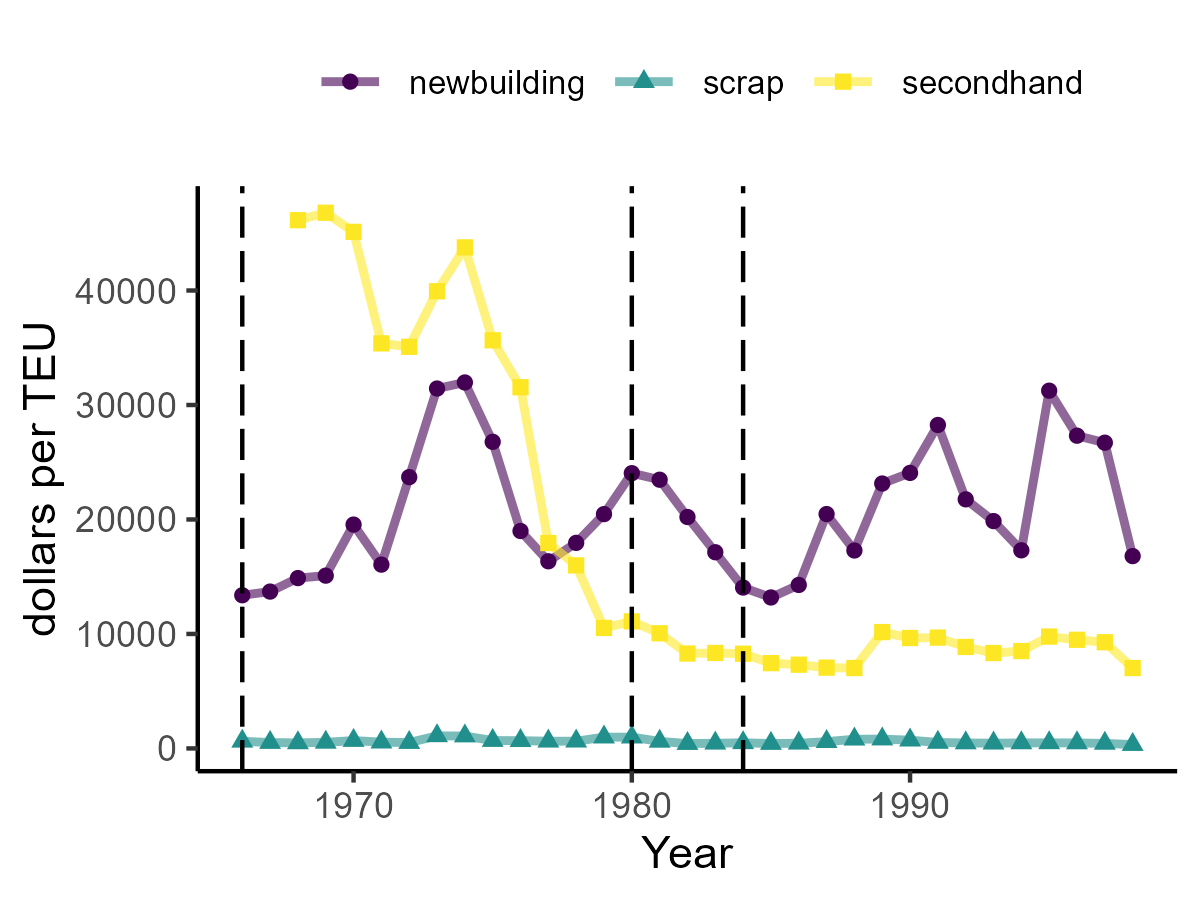}
\caption{Trends in industry-year-level newbuilding, secondhand, and scrap prices.}
\label{fg:price_newbuilding_secondhand_scrap}
\end{center}
\footnotesize
The data span 33 years (1966-1998) for the entire container shipping industry. The data cover newbuilding and scrap prices between 1966 and 1998 and secondhand prices between 1968 and 1998. All prices were adjusted according to the CPI in the U.S. in 1995. 
\end{figure}

Figure \ref{fg:price_newbuilding_secondhand_scrap} illustrates the transitions of newbuilding, secondhand, and scrap prices. First, we observe four peaks in the newbuilding price in 1974, 1980, 1987, and 1991, indicating that the newbuilding price peaks at similar times to the shipping price. Surprisingly, the shipping price, after adjusting CPI, did not decrease significantly, which means that the trend of the newbuilding price did not decrease unlike the shipping price. This is a new finding because it is commonly known that vessels themselves were expensive in the 1970s and the shipbuilding price per TEU decreased in the 1990s due to the gradual increase in the size of ships. We find that the increasing CPI corresponds with a seemingly decreasing shipbuilding price in the nominal sense. 

Second, the secondhand price sharply decreased between 1974 and 1979. However, after 1980, it remained relatively stable compared to shipping and newbuilding prices. In the 1970s, when full container vessels started operating on liner routes, the number of vessels was insignificant, leading to a small number of vessels traded in the secondhand market, which resulted in more fluctuation in secondhand prices. However, since the 1980s, the number of vessels traded in the secondhand market has increased, leading to more stable secondhand freight rates and a more significant correlation with new building prices.

Third, the scrap price remained stable, without significant fluctuations relative to other price variables. Moreover, the scrap price level was smaller than that of the newbuilding and secondhand prices because the scrap price is linked to the input price, such as the steel price. Thus, the trend in scrap price reflects the dynamics of the input price.

\subsection{Summary statistics}\label{subsec:summary_statistics}

\subsubsection{Route-year-level shipping price and quantity data}

\begin{table}[!htbp]
  \begin{center}
      \caption{Summary statistics.}
      \label{tb:summary_statistics} 
      \subfloat[Route-year-level variables (1966-2008)]{
\begin{tabular}[t]{lrrrrr}
\toprule
  & N & mean & sd & min & max\\
\midrule
Price (\$ per TEU): $P_{rt}$ & 240 & 2105.35 & 1250.84 & 561.86 & 6654.79\\
Quantity (1 mil TEU): $Q_{rt}$ & 240 & 2.34 & 2.75 & 0.00 & 17.70\\
\bottomrule
\end{tabular}
}\\
      \subfloat[Industry-year-level variables: $(P_{t}^{new},P_{t}^{scrap})$ (1966-1998) and $P_{t}^{second}$ 1968-1998)]{
\begin{tabular}[t]{lrrrrr}
\toprule
  & N & mean & sd & min & max\\
\midrule
Newbuilding price (\$ per TEU): $P_{t}^{new}$ & 33 & 20633.82 & 5555.70 & 13186.28 & 31966.62\\
Secondhand price (\$ per TEU): $P_{t}^{second}$ & 31 & 18371.82 & 14502.46 & 7020.96 & 46792.94\\
Scrap price (\$ per TEU): $P_{t}^{scrap}$ & 33 & 609.97 & 196.54 & 334.25 & 1098.38\\
\bottomrule
\end{tabular}
}
  \end{center}\footnotesize
  Note: Panel (a) of the six routes covers a data span of 44 years (1966-2009). The Transatlantic routes opened in 1966, Transpacific routes opened in 1967, and Asia-Europe routes opened in 1971. The shipping prices before 1992 in Panel(a) refer to conference prices, while the shipping prices after 1993 are unified prices based on conference and non-conference prices. The difference between them vanished because of the Shipping Act of 1984. The latter are standard data that are often used in the literature such as \cite{jeon2022learning}. Panel (b) of the container shipping industry covers a data span of 33 years (1966-1998). The data cover newbuilding and scrap prices between 1966 and 1998, and secondhand prices between 1968 and 1998. All prices were adjusted according to the CPI in the U.S. in 1995. 
\end{table} 

Panel (a) in Table \ref{tb:summary_statistics} presents the summary statistics of the route-year-level variables for the shipping price $P_{rt}$ and quantity $Q_{rt}$. The data cover the Transatlantic routes that opened in 1966, the Transpacific routes that opened in 1967, and the Asia-Europe routes that opened in 1971.\footnote{For Table \ref{tb:summary_statistics}, the shipping prices before 1992 refer to conference prices, while the shipping prices after 1993 are unified prices based on conference and non-conference prices, while the difference between them vanished in \textit{the Containerization International Yearbook}. Thus, we observe the full history of the conference market.} On average, the shipping price was \$2105.35 per TEU and the shipping quantity was 2.34 million TEUs. 

\subsubsection{Industry-year-level newbuilding, secondhand, and scrap prices data}

Panel (b) in Table \ref{tb:summary_statistics} shows the summary statistics of industry-year-level variables for the newbuilding price $P_{t}^{new}$, scrap price $P_{t}^{scrap}$ between 1966 and 1998, and secondhand price $P_{t}^{second}$ between 1968 and 1998. The average newbuilding price per TEU was \$ 20633.82. The average secondhand price per TEU was \$18371.82, which was less than the newbuilding price after 1971. The average scrap price per TEU was \$609.97.

\subsection{Institutional details}\label{subsec:institutional_details}

Here, we introduce the institutional background of the shipping conference, the early history of the container shipping industry, and two key events triggering the change in competition regime.

Table \ref{tb:industry_history} summarizes the historical events of the liner shipping industry before and after global containerization.
\begin{table}[ht!]
    \caption{The historical background of the liner shipping industry.}
    \label{tb:industry_history}
    \centering\scriptsize{}
    \begin{tabular}{cll}
      Timing / Period & Event & Related Study\\\hline
      1875 & the first liner shipping conference (The U.K.-Calcutta) established & \cite{morton1997entry}\\
       &  & \cite{podolny1999social}\\
      1916 & \textbf{The Shipping Act of 1916} prohibited deferred rebate &\\
      1918 & The end of WW1 & \cite{deltas1999american}\\
      1936 & U.S. Maritime Commission established &\\
      1945 & The end of WW2 & \\
      1956 & The world's first container ship sailed & \\
      1961 & \textbf{The 1961 Amendment} enacted & \\
      & \textbf{Federal Maritime Commission (FMC)} established & \\
      & Technical Committee on Cargo Containers established (ISO/TC104) &\\
      1964 & ISO adopted seven sizes of containers as ISO standards &\\
      \hline
      1966 & The first Full-container (FC) ship sailed the Transatlantic route &\\
      1967 & The first FC ship sailed the Transpacific route &\\
      1971 & The first FC ship sailed the Asia-Europe route& \\
      1973 & Oil shock &\\
      1974 & The UNCTAD Code of Conduct for Liner Conferences adopted &\cite{fox1992empirical,fox1995some}\\
      1980 & \textbf{Withdrawal of Sea-Land from shipping conferences}&\cite{sjostrom1989collusion}\\
      1983. April&The UNCTAD Code of Conduct for Liner Conferences accomplished &\\
      1984. June & \textbf{The Shipping Act of 1984} in the U.S. enacted, conference breakdown &\cite{clyde1995effectiveness,clyde1998market}\\
      & &\cite{wilson1991some}\\
      & &\cite{pirrong1992application}\\
      1985.Sep & Yen appreciation due to the Plaza Accord&\\\hline
      1991 & Strategic alliance boom started &\\
      1998. Oct  & \textbf{The Ocean Shipping Reform Act (OSRA) of 1998}, effective in 1999 &\cite{reitzes2002rolling}\\
      &&\cite{fusillo2006some,fusillo2013stability}\\\hline
    \end{tabular}
    \begin{tablenotes}
\item[a]\textit{Note:} This table is based on unified survey papers such as \cite{sjostrom2004ocean,sjostrom2013competition} and \cite{martin2012market} and specific papers \citep{clyde1995effectiveness,clyde1998market}. Industry and legislative changes after 1990 are beyond the scope of this study. For reference, see \cite{reitzes2002rolling}, which states that ``OSRA alters the role of the Federal Maritime Commission as a cartel enforcer. Under the Shipping Act of 1984, all carriers, both conference carriers, and independent carriers, had to file their tariffs with the FMC. The FMC then policed these rates, issuing fines to carriers that engaged in secret discounting, known as ``rebating." Under the OSRA, carriers are obliged to make their freight rates publicly available, but the FMC's enforcement obligations are eliminated. OSRA's elimination of tariff-filing requirements and rate enforcement by the Federal Maritime Commission raises the cost of monitoring their members' pricing activities. (page 56)."
   \end{tablenotes}
\end{table}

\subsubsection{Shipping conferences}\label{subsec:shipping_conference}
The world's first shipping conference, the ``Calcutta Conference," was formed in 1875 on the route between England and Calcutta to establish a unified freight rate.\footnote{In this subsection, we refer to \textit{The Actual State of Competition in Oceangoing Shipping and Problems with Competition Policy} (``Gaikoukaiun no Kyousoujittai To Kyousouseisakujou no Mondaiten ni Tsuite," in Japanese) reported by \cite{gaikoukaiun_no_kyousoujittai2006} of the Japan Fair Trade Commission, \cite{branch2013maritime}, and \cite{sjostrom2013competition}. In particular, \cite{sjostrom1989collusion} discusses the economic insights of the shipping conferences from an historical viewpoint. \cite{morton1997entry} provide detailed historical evidence on British shipping industry before 1900.} In 1879, the Chinese Alliance, the ancestor of the Far East Freight Conference, was formed on routes between Asia and Europe.\footnote{\cite{sjostrom2004ocean} conducted a survey on shipping conferences, and pointed out that systems similar to shipping conferences had existed in Atlantic shipping and British coastal shipping before the Calcutta Conference.} The shipping conferences appeared 15 years before the enactment of the Sherman Antitrust Act in 1890, one of the first competition laws in the modern world, so cartels themselves were not illegal at that time. Table \ref{tb:industry_history} summarizes the history of the liner shipping industry before 2000, with related studies and key legislative changes.

\paragraph{Internal mechanism to conference firms.}
The internal mechanism of conference firms is mainly aimed at market stabilization by controlling entry via excess capacity \citep{fusillo2003excess}, predatory pricing \citep{morton1997entry,podolny1999social}, price discrimination \citep{fox1992empirical,clyde1998market}, and loyalty contracts \citep{marin2003exclusive}, among other things. To achieve this, shipping conferences agreed on various matters, in addition to freight rates. The content of the agreements covered: (1) alternatives to suppress freight rate competition among member shipping companies, (2) alternatives to prevent shippers from moving to non-conference shipping companies, and (3) alternatives to directly exclude non-conference vessels.

To avoid freight rate competition, rate agreements and vessel allocation agreements were concluded among the member shipping companies. Rate agreements are signed by members to agree on rates for each product and to update the rates jointly. Vessel allocation agreements adjust the amount of tonnage to be allocated, the number of voyages, ports of call, operation schedules, and cargo to be loaded. These features were modeled as price and quantity-fixing cartels.\footnote{For example, \cite{clyde1998market} model the market with the shipping conference as a collusion equilibrium.}

\paragraph{External mechanism for non-conference firms.}
The conference tariff rates, or freight rates, determined by these agreements had been publicly noticed, and no entry restrictions imposed non-conference container shipping market. The market was, however, subject to competition from non-conference vessels and new entrants. Consequently, shipping conferences introduced the Dual Rate System, the Fidelity Rebate System, and the Differed Rebate System to ensure effectiveness of freight rate agreements and prevent shippers from flowing to non-conference shipping companies that offer lower rates than the conferences' rates.\footnote{In the Dual Rate System, a shipper and a shipping conference conclude an exclusive patronage contract/loyalty agreement and provide transportation service for the specific route. The contract rate is lower than the spot rate on the condition that the shipper uses only conference member carriers' service within a specific contract period. The Fidelity Rebate refunds a portion of the freight if the shipper uses only the conference carriers within a specified period (4-6 months). The Differed Rebate System is also an incentive to use conference carriers. Under the system, if a shipper had used only members' services for a specified refundable period (4-6 months), and if it does not use any non-conference members' service for the deferment period following the refundable period, a certain amount of money is refunded upon the shipper's request. The refund amount was usually around 10\% of the freight. \cite{fox1992empirical} uses the U.S. port pair-level data in 1977 to examine the effect of the dual rate contract and consumer loyalty.} These entry deterrents promoted the stability of freight rates and liner services.\footnote{For instance, \cite{marin2003exclusive} examine the economic effects of exclusive contracts of ocean shipping cartels during the 1950s between firms and the ultimate consumers of their product. They record that, ``During the congressional investigations of shipping conferences in the late 1950s and early 1960s documents obtained from an ocean carrier contained an admission that ‘the entire contract system is a fighting measure to get rid of outside competition' " (p.198).}


Conferences also utilized ``fighting ships," which are vessels that temporarily put into service at a similar schedule as non-conference shipping companies and at lower rates.\footnote{See \cite{marshall2014economics} (page 148) and \cite{harrington2018rent} for reference to put the strategy of shipping conferences in general cartel literature. \cite{harrington2018rent} classify the general response of cartels to the expansion of noncartel supply into four strategies: takeover,
starvation, coercion, and bribery. The fighting ships are classified into coercion strategy.} Fighting ships were used to force non-conference shipping companies to leave routes, and all members shared the associated operational losses.

Since the 1960s, the nature of shipping alliances have changed. Technological innovation centered on containerization, and pro-competitive amendments to shipping laws in the United States, particularly, have had an impact on the functioning of shipping alliances and market competition in the shipping market.

\subsubsection{The inception of the container shipping industry}\label{subsec:inception_of_container_shipping_industry}

The history of container shipping began with Malcolm P. McLean, the founder of the U.S. Land Transportation Company Sea-Land Service.\footnote{\cite{bernhofen2016estimating} and \cite{rua2014diffusion} explain the detailed history of the container shipping industry from the viewpoint of the global trade and country-level development. \cite{levinson2016box} provides an overview of the industry with anecdotal and qualitative evidence.} The world's first container ship sailed from the Port of Newark, New Jersey, to the Port of Houston, Texas in 1956. The first international container ship was employed by Sea-Land Service for the Transatlantic Route in 1966, and for the Transpacific route in 1967. However, for the Asia-Europe route, the first international semi-container ship, Cornelia Maersk, was employed in 1967. The first international full-container ship, Kamakura-maru, was delayed in 1971.

Global containerization induced several market changes between 1966 and 1990, including transforming the cost structure, lowering barriers to entry, stimulating the rise of non-conference shipping companies in developing countries, shifting from ``closed" conferences to ``open" conferences, and forming a consortium. Appendix \ref{sec:details_of_global_containerization} provides further details on each of these changes.

\subsubsection{Withdrawal of Sea-Land from shipping conferences in 1980}\label{subsec:withdrawal_of_sea-land_from_cartel_in_1980}
In the late 1970s, the increase in competition for Transpacific routes led to pressure on shipping rates. To address this, Sea-Land introduced eight SL-7 high-speed container vessels with a speed of 33 knots. However, the high operation costs of these vessels negatively impacted the company's profitability. Consequently, Sea-Land withdrew from shipping conferences to avoid tonnage-based cargo allocation imposed by the conferences and secure profitability. Former executives of a Japanese shipping company stated that Sea-Land had to increase loaded cargo but could not achieve its goals without leaving the conference that set freight rates and that it felt it had no choice but to become a non-alliance carrier and establish its own freight rate to retain customers \citep{JapanMaritimeDaily2006}.

After Sea-Land's withdrawal from shipping conferences, shippers with contracts with conference carriers hesitated paying the penalty to switch to Sea-Land under the conference's Dual Rate System. Sea-Land focused on import shippers and intermodal cargoes not covered by the double-freight rate system. The company also paid the freight cost of returning inland containers to the port instead of passing them on to shippers, decreasing freight rates.

\subsubsection{The Shipping Act of 1984 in the U.S.}\label{subsec:shipping_act_of_1984}

In June 1984, the United States enacted the new Shipping Act as part of the Reagan administration's deregulation policy to allow member carriers to make individual agreements with shippers on freight rates and services, and to unbind shippers from shipping conferences to enable them to make more appropriate choices in shipping companies. This drastically changed the competition regime.\footnote{\cite{wilson1991some} provide anecdotal evidence and a case study of the effect of the Shipping Act of 1984.}

The Shipping Act of 1984 included the right to Independent Action (IA), allowing member firms to define freight rates or services that deviate from the conference tariff rates and guaranteeing of members carriers in the shipping conferences to set their own rates or services.

It required conferences to allow the right to form service contracts (SCs), which are contracts in which the shipper commits in advance to load a specific quantity or more of cargo to the shipping company during a specific period. Under this contract, the shipping company reserves the space necessary to carry cargo and applies discounted freight rates\footnote{The minimum number of containers promised by the shipper to the shipping company is called the MQC (Minimum Quantity Commitment).}.

The act explicitly prohibited the Dual Rate System and the shipping conference lost its binding power over shippers, and individual shipping companies frequently exercised their right to independent action, which encouraged competition and led to a significant decline in freight rates on U.S.-related routes.\footnote{\cite{JMC2008} pointed out that the Dual Rate System was the most effective way in which shipping companies kept their shippers when the conference system was functioning.}

\paragraph{Stabilization agreement.}
Containerization and the Shipping Act of 1984, weakened the price-binding power of shipping conferences by increasing contracts based on IAs and SCs. Shipping companies thus set their freight rates, leading to intense price competition and sharp drops in Transpacific freight rates. To stabilize shipping routes, conference shipping companies formed a ``stabilization agreement" inviting most shipping companies, including non-conference ones.

Agreements shared information on supply and demand trends and agreed guidelines for rate restoration and surcharges, but binding power on rates. The Transpacific Stabilization Agreement (TSA) was formed in 1989 by 13 shipping companies to stabilize shipping routes from Asia to North America.\footnote{TSA was dissolved in 2018.}



\section{Interviews}\label{sec:interview}

In this section, we provide interview-based evidence on the consistency of our recovered data on container freight rates with the historical experience of industry experts, Akimitsu Ashida and Hiroyuki Sato.\footnote{We obtained additional interview-based evidence. Mikio Tasaka, who belonged to the Nittsu Group and worked in the U.S. in the 1980s, stated that there was no discrepancy in the development of freight rates. In addition, Carolyn Almquist experienced pricing and conference division in the container shipping business at American President Lines, stated that provide a reasonable basis for analysis. Professor Yutaka Yamamoto (University of Nagasaki), who had 20 years of experience in the container shipping business at American President Lines, stated that the trend was generally reasonable. We also received responses from Kwon Oh In who worked for 40 years at Korea Maritime Transport Company, a major Korean container shipping company. He said that our graphs seemed convincing as a general trend of the container shipping industry during that period.}

\subsection{Akimitsu Ashida, an ex-chairperson of Mitsui O.S.K Lines}

Concerning the increase in freight rates in Transatlantic and Asia-Europe routes in the 1980s, Mr. Akimitsu Ashida, former chairman of Mitsui O.S.K. Lines (MOL), answered about the situation on liner routes. He served as the company's European Division Manager from 1985-86 and responded to our e-mail inquiry on February 28, 2022. We asked him about his thoughts and related memory on the figures and tables.\\


\textbf{Regarding Shipping Quantity in Asia-Europe routes} \\
Ashida: Unlike Transpacific routes (where U.S. Customs publishes data), European routes did not have a system whereby customs authorities published statistics on container transport volumes. Therefore, the FEFC did not have to compile and notify its members of actual container transport volumes. Consequently, it is not easy to find the data even today. 

\textbf{Regarding Freight Rates in Asia-Europe routes} \\
Ashida: Freight rates were rising regularly in nominal terms. One of the reasons was that shipping conferences had relatively strong power on Asia-Europe routes. In addition, surcharges such as the Bunker Adjustment Factor (BAF) and Currency Adjustment Factor (CAF) were collected without fail. Under these circumstances, the Plaza Accord of 1985 raised the yen-dollar exchange rate from 240 yen to approximately 140 yen. This caused dollar-based freight rates to rise sharply, especially for cargoes originating in Japan. Non-Japanese carriers earned even higher profits because their yen-based costs were relatively small. 

\textbf{The Reason Why the Shipping Conferences Were Still Functioning on the Asia-Europe Routes}\\
Ashida: One reason is that there were no powerful non-conference member shipping lines. At that time, Japanese shippers avoided non-conference members when exporting cargo from Japan. The exports accounted for 50 \% of all cargo shipped from Asia. Therefore, the export cargo was mainly assigned to conference member firms. This trend was disrupted by the introduction of large newly built vessels by non-conference members from 1990 onward.

I recall that about 35 \% of cargo on the Asia-Europe routes was handled through forwarders, unlike the Transpacific routes. Furthermore, the Asia-Europe routes were less active in discount negotiations than the Transpacific routes because the inland transport distances were shorter and lower ocean freight rates would reduce the profit for the company.

\subsection{Hiroyuki Sato, an ex-vice president of Mitsui O.S.K Lines}

We interviewed Mr. Hiroyuki Sato, former vice president of MOL, about the shipping conferences in the Transpacific routes and the situation in liner routes in the 1970s and the 1980s. He had been in charge of sales for Asia-Europe service routes from 1969 and Transpacific service routes from 1974. We conducted onsite interview with Mr. Sato on November 17, 2021 and our e-mail inquiries were responded by him on December 15, 2021 and February 28, 2022. We asked him about his thoughts on the figures and tables used in this study.\\



\textbf{Regarding Freight Rate}\\
Sato: In the 1970s, containerization was progressing on all routes, but freight rates were determined based on weight or measured in the same way as conventional vessels, instead of box rates (rates per container). There was also an arrangement called ``minimum revenue per container." In addition, a few percent of discounts were applied to long-term contract shippers using a double freight contract with the shipper. 

In the 1980s, there was a shift to box rates; the enactment of the Shipping Act of 1984 was a significant reason for this shift. The period after the Act was when space charters were no longer viable, leading to the formation of alliances in the 1990s.

\section{Results of Structural Breaks Test}\label{sec:structural_break_test}

We test multiple unknown structural breaks for each route using \cite{bai1998estimating,bai2003computation}, to confirm anecdotal evidence of container crises between the 1970s and the 1980s, documented in \cite{broeze2002globalisation}, and to find when the container crisis started and how long it lasted. Methodologically, we follow \cite{bai2003computation} to address the problem of estimating the break dates and the number of breaks and present an efficient algorithm.\footnote{The most relevant paper is \cite{fan2016analysis}, which applies \cite{bai2003computation}'s method to the semi-annual data on the newbuilding price index, the time charter rate index, and the second-hand price index for each ship size (i.e., Feeder, Feedermax, Handy, Sub-Panamax, and Panamax) between October 1996 and July 2013. They focus on the unknown structural breaks in the relationship between the three abovementioned global-level indices, whereas we are interested in the unknown structural breaks in the route-level container freight rate corresponding with competition regime changes.} 

We consider a multiple linear regression with $k$ breaks ($k + 1$ regimes) as follows:
\begin{align}
    y_{t}=z_{t}^{\prime} \delta_{j}+u_{t} \quad t=T_{j-1}+1, \ldots, T_{j}\label{eq:bai_and_perron}
\end{align}
for $j=1, \ldots, k+1$. In this model, $y_{t}$ is the targeted price at time $t$, $z_{t}(q \times 1)$ is a vector of $q$-dimensional covariates, $\delta_{j}(j=1, \ldots, k+1)$ is the corresponding vector of coefficients, and $u_{t}$ is the disturbance at time $t$. We allow serial correlation in the errors and variances of the residuals across segments. The indices $\left(T_{1}, \ldots, T_{k}\right)$, that is., the breakpoints, are explicitly treated as unknown (we use the convention that $T_{0}=0$ and $T_{k+1}=T$ ). 

The purpose of this section is to estimate the unknown regression coefficients together with the breakpoints and the number, therefore we specify $z_{t}=1$ as the simplest case. \cite{bai2003computation} suggested that the trimming parameter $\varepsilon=h/T=0.10$ holds where $h$ is the minimum distance between each break. Following \cite{bai2003computation}, we assume a minimum length equal to 10\% of the sample size, given that we allow for a maximum of four breaks. Our data specify $T=40$ (i.e., yearly data between 1966-2007), thus we impose $h=4$ for the route-year-level analysis of the shipping price. For consistency, we also impose $h=4$ for the industry-year-level analysis of newbuilding, secondhand, and scrap prices.

\subsection{Route-year-level shipping price data}

\begin{figure}[!ht]
\centering
\begin{minipage}[b]{0.3\linewidth}
  \includegraphics[keepaspectratio, scale=0.31]
  {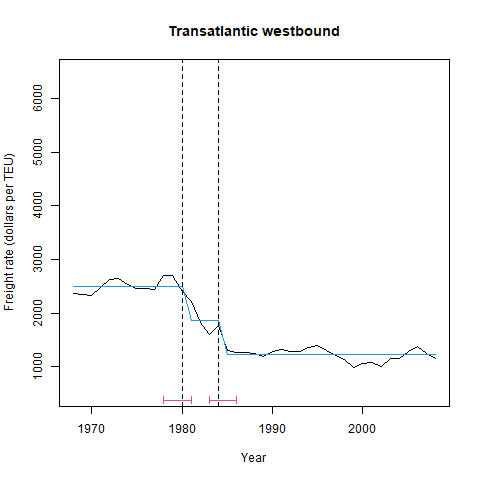}
  \end{minipage}
  \begin{minipage}[b]{0.3\linewidth}
  \includegraphics[keepaspectratio, scale=0.31]
  {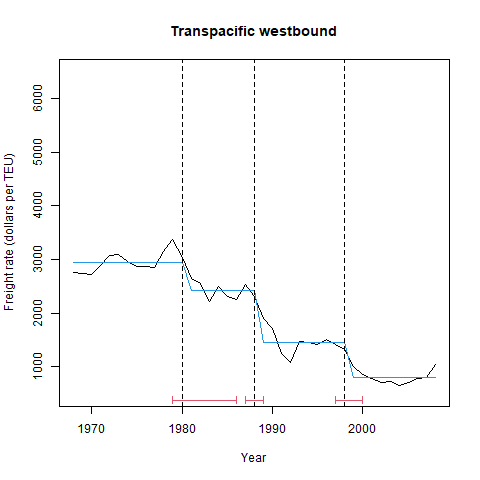}
  \end{minipage}
  \begin{minipage}[b]{0.3\linewidth}
  \includegraphics[keepaspectratio, scale=0.31]
  {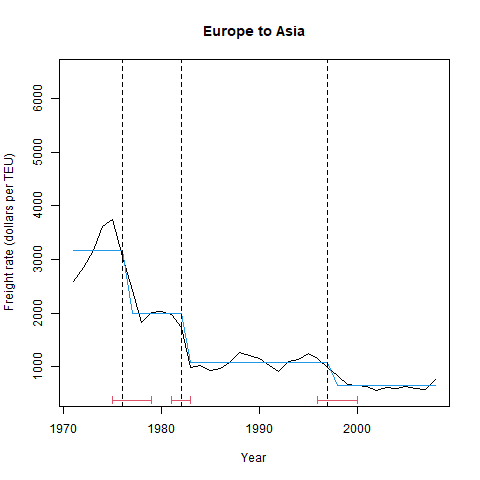}
  \end{minipage}\\
\begin{minipage}[b]{0.3\linewidth}
  \includegraphics[keepaspectratio, scale=0.31]
  {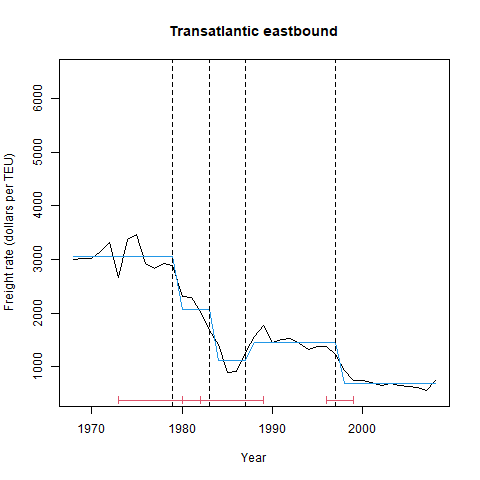}
  \end{minipage}
  \begin{minipage}[b]{0.3\linewidth}
  \includegraphics[keepaspectratio, scale=0.31]
  {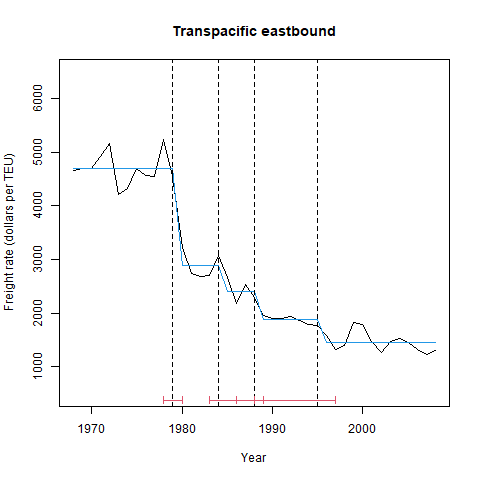}
  \end{minipage}
  \begin{minipage}[b]{0.3\linewidth}
  \includegraphics[keepaspectratio, scale=0.31]
  {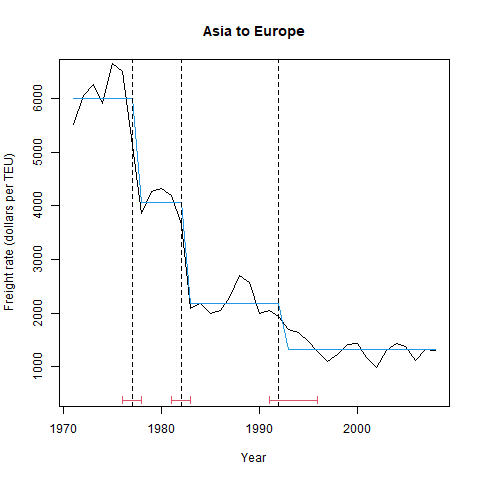}
  \end{minipage}
  {\scriptsize{}
  
\begin{tabular}[t]{llllllllll}
\toprule
  & BIC: $m$ = 0 & $m$ = 1 & $m$ = 2 & $m$ = 3 & $m$ = 4 & $m$ = 5 & $m$ = 6 & $m$ = 7 & $m$ = 8\\
\midrule
Transatlantic westbound & 647.86 & 555.88 & 538.08 & 538.10 & 538.89 & 538.95 & 545.65 & 552.91 & 560.26\\
Transatlantic eastbound & 685.71 & 620.56 & 596.29 & 582.23 & 581.03 & 584.06 & 587.28 & 592.78 & 598.48\\
Transpacific westbound & 679.59 & 609.32 & 592.63 & 568.14 & 568.96 & 573.59 & 579.50 & 586.58 & 593.87\\
Transpacific eastbound & 714.70 & 640.64 & 599.18 & 592.70 & 592.22 & 595.82 & 600.01 & 606.82 & 614.56\\
Asia to Europe & 684.17 & 620.42 & 593.91 & 567.82 & 571.46 & 576.96 & 584.01 & 591.20 & 627.68\\
Europe to Asia & 631.38 & 581.33 & 553.76 & 539.17 & 545.08 & 551.75 & 558.62 & 566.59 & 588.79\\
\bottomrule
\end{tabular}

\caption{The estimated breakpoints and 95 \% confidence intervals with BIC.}
\label{fg:structural_change_transatlantic_eastbound_before_2008}
\begin{tablenotes}
\item[a]Note: Each segment shows the estimated breakpoints and the 95\% confidence intervals for $\hat{T}_i$ for $i=1,\cdots,4$. Following \cite{bai2003computation}, the number of breakpoints is selected by minimizing BIC, as shown in the bottom table. The estimated parameters $\hat{\delta}_j$ and standard errors for all $j=1,\cdots,k+1$ for all possible numbers of breakpoints are omitted. We use the sample between 1968 and 2008 for the Transatlantic and Transpacific routes and the sample between 1971 and 2008 for the Asia-Europe routes. We used the \texttt{strucchange} R-package developed by \cite{zeileis2002strucchange}.
   \end{tablenotes}
   }
\end{figure}

Figure \ref{fg:structural_change_transatlantic_eastbound_before_2008} illustrates the estimated breakpoints $\hat{T}_j$ and 95\% confidence intervals for route-year-level analysis of shipping freight rates. The bottom table shows BIC for all possible number of breaks under $h=4$, where the minimizer of the BIC is the optimal break number $\hat{k}$. First, we find that the 95\% confidence intervals of the structural break cover the period between 1979 and 1980 on the U.S. routes (left and center panels), whereas the intervals cover the period between 1976 and 1978 on the non-U.S. routes (right panels). This indicates that the container crisis occurred on all six routes by 1980, triggered by the withdrawal of Sea-Land, although non-U.S. routes reacted earlier than U.S. routes. The reason for the response in non-U.S. routes may be related to the reopening of the Suez Canal in 1976, which increased the supply of container shipping services \citep{jsme2022}.

Second, we find remarkable differences between U.S. routes and non-U.S. routes between 1984 and 1990. The 95\% confidence intervals of the structural break cover some period between 1984-1990 on the U.S. routes, whereas there were no breaks on the non-US routes in the period. This implies that the enactment of the Shipping Act of 1984 generated some structural breaks in the treatment group — the U.S. routes. Surprisingly, we also find that the container crisis lasted heterogeneously, even along the U.S. routes. For example, Transpacific routes seemed to recover in 1986, whereas Transatlantic routes remained sour until 1990. This result may support the fact that the supply-demand relationship on the Transpacific and Transatlantic routes changed differently in the late 1980s. The substantial change in exchange rates after the Plaza Accord in 1985 led to a recovery in the demand for transportation on the Transpacific route. However, no such situation was observed between Europe and the U.S.

\subsection{Industry-year-level newbuilding, secondhand, and scrap prices data}

\begin{figure}[!ht]
\centering
\begin{minipage}[b]{0.3\linewidth}
  \includegraphics[keepaspectratio, scale=0.31]
  {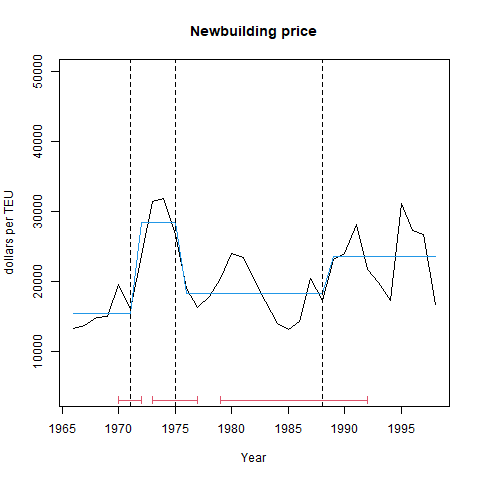}
  \end{minipage}
  \begin{minipage}[b]{0.3\linewidth}
  \includegraphics[keepaspectratio, scale=0.31]
  {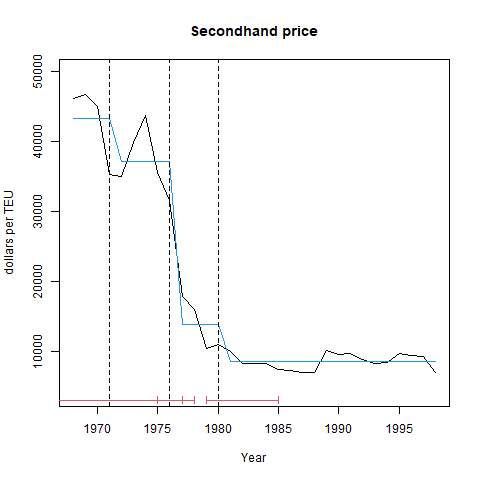}
  \end{minipage}
  \begin{minipage}[b]{0.3\linewidth}
  \includegraphics[keepaspectratio, scale=0.31]
  {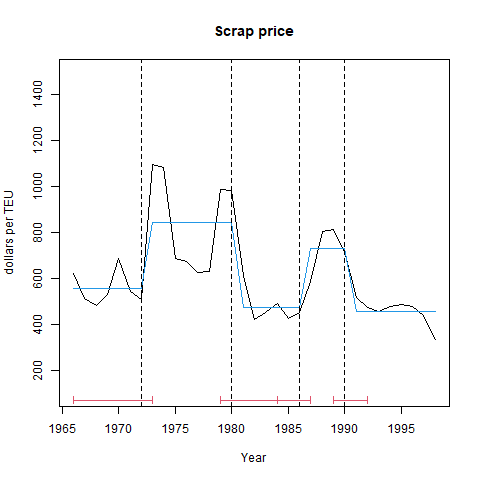}
  \end{minipage}
  {\scriptsize{}
  
\begin{tabular}[t]{lllllllll}
\toprule
  & BIC ($m$ = 0) & 1 & 2 & 3 & 4 & 5 & 6 & 7\\
\midrule
Newbuilding & 668.72 & 668.35 & 665.16 & 661.39 & 663.70 & 667.99 & 674.65 & 694.26\\
Secondhand & 687.91 & 610.21 & 609.25 & 606.64 & 612.79 & 619.02 & 625.65 & -\\
Scrap & 448.16 & 446.64 & 440.37 & 443.81 & 440.36 & 446.38 & 452.99 & 470.14\\
\bottomrule
\end{tabular}

\caption{The estimated breakpoints and 95 \% confidence intervals with BIC.}
\label{fg:structural_change_scrap_price_before_1998}
\begin{tablenotes}
\item[a]Note: Each segment shows the estimated breakpoints and the 95\% confidence intervals for $\hat{T}_i$ for $i=1,\cdots,4$ for each route. Following \cite{bai2003computation}, the number of breakpoints is selected by minimizing BIC, as shown in the bottom table. The estimated parameters $\hat{\delta}_j$ and the standard errors for all $j=1,\cdots,k+1$ for all possible numbers of break points are omitted. We use samples from 1968 to 1998 and the \texttt{strucchange} R-package developed by \cite{zeileis2002strucchange}. 
   \end{tablenotes}
   }
\end{figure}

Figure \ref{fg:structural_change_scrap_price_before_1998} illustrates the estimated breakpoints $\hat{T}_j$ and 95\% confidence intervals for industry-level analysis of newbuilding, secondhand, and scrap prices. First, we find that 95\% confidence intervals of the structural break cover the period between 1974 and 1976 for newbuilding and secondhand prices. These results correspond with the shipping price trend in the Asia-Europe market. Second, downward structural breaks in prices were not detected between 1984 and 1990. These results imply that, unlike route-level shipping prices, the container crisis did not significantly decrease industry-level prices in the shipbuilding market. We interpret these differences as naturally capturing the differences between shipping markets with shipping conferences and those without cartel groups.

\section{Practical Implications, Discussion, and Future Research}

\subsection{Practical implications}
Our study makes an important contribution to the development of consistent data on the history of the container shipping industry and policy discussions for practitioners. The data contribution enables practitioners to obtain empirical knowledge on the main container transport markets, as well as to develop a methodology to construct container market trends on other routes, where information such as freight rates and vessel volumes deployed is only partially available. For example, container transport market trends on routes between the Indian subcontinent, Africa and Europe, South East Asia and the U.S. are important information for manufacturing companies to establish their supply chains and respond to specific policies, but the available information is fragmented. Thus, our data contribution sheds light on the calculation of partially known logistics costs for shippers.

\subsection{Discussion}
We summarize the potential concerns for the data and structural break tests. 
First, our imputation approach is ad hoc, which potentially makes the data sensitive to other imputation approaches or assumptions. 
However, we believe that our approach is the best alternative among feasible approaches because other data-driven imputation approaches require a large sample size, several overlapping years, and the no-structural-break assumption of the data, which our data does not satisfy (discussed in Section \ref{subsec:institutional_details} and Appendix \ref{sec:data_construction}). 
Second, we may face a small sample problem for the structural breaks tests because we use route-year-level data between 1968 and 2008. 
If route-quarter-level data for a longer period can be consistently constructed, the results of the structural breaks may change. 

Regarding the possibility of combining our data with data from 2009 and beyond, we have a couple of reasons for not doing so. First, our main data source, Review of Maritime Transport, only provides route-level price data up until 2009. 
Second, there are some challenges with merging our data with Drewry's data for a fee since their price data is based on a different observation unit than ours. Drewry's data includes monthly port-to-port level price data, while our data is route-year-level.
Third, the industry experienced significant changes around 2009, such as the Global Financial Crisis and the Repeal of the EU's competition law exemption for shipping conferences in October 2008. These industry changes not only make it inappropriate to simply merge data from 2009 and beyond with our existing data, but they can also contaminate the results of structural break tests.
These concerns are beyond the scope of this study and are left for future research.

\subsection{Future research}

Our data are useful for understanding the related historical policy debates in the container shipping industry. 
In ongoing research, one of the authors merge our data with the firm-market-level shipbuilding data between 1966 and 1990 ; then construct a dynamic structural model of firms' entry, exit, and shipbuilding investment decisions in homogenous good markets under collusive and perfect competition regimes, which are exogenously determined by the enactment of the Shipping Act of 1984. 
Understanding the effect of shipping conferences based on the structural model makes the specific features of the inner allocation mechanism of shipping conferences contributable to the general literature on competition law and industrial policy which have been investigated in the shipbuilding and shipping industry \citep{kalouptsidi2017res,barwick2019china}.
Thus, our data sheds light on the historical questions of future research in the container shipping industry.

Our data are also useful for understanding the current policy debates in the container shipping industry. 
For example, we merge our data with post-1990 firm-market-level shipbuilding data to evaluate the effects of EU competition law exemptions (by Regulation (EEC) No 4056/86) for liner shipping firms, which terminated in 2008, on merger waves after 1990. 
\cite{jeon2022learning} uses data from 2006 to 2014 and conducts counterfactuals with respect to competition and industry consolidation by simulating the industry under a multi-plant monopolist and a merger of the top two firms. \cite{kalouptsidi2017res} and \cite{barwick2019china} investigated the Asian shipbuilding market in the 2000s to study China’s expansion. 
Thus, our data fill the gap between prior studies focusing on the 2000s and the 2010s and future research examining the 1980s and the 1990s.

Our data provides a historical benchmark of shipping prices for updated issues in the container shipping industry. First, our data enables us to make quantitative comparisons between price increase during Covid period and historical prices that were under active shipping conferences. 
For instance, during the Covid period, the highest peak of shipping prices per TEU from Shanghai to Los Angeles was recorded in February 2022, adjusted according to the CPI in the U.S. in 1995, and amounted to \$4462. 
This level is comparable to the highest peak of cartel prices observed in the Transpacific eastbound route, as illustrated in Figure \ref{fg:container_freight_rate_and_shipping_quantity_each_route}.
This will offer valuable insights into how prices have evolved over time, as well as how significant the shipping price increase during Covid was. 
Second, our 1966–2009 data provides a helpful benchmark for predicting the impacts of significant industry changes that are expected to occur, such as the updates to the EU competition law exemption for consortium in 2024 and the breaking of the 2M alliance between Mediterranean Shipping Company (MSC) and Maersk in 2025. 
By using our data, we can gain a better understanding of the potential effects of these changes on shipping prices.

\section{Conclusion}\label{sec:conclusion}
This study provides a new unified panel dataset of route-year-level freight rate and shipping quantities for the six major routes and industry-year-level newbuilding, secondhand, and scrap prices from 1966 (beginning of the industry) to 2009. The data provide a fundamental basis for understanding the container shipping industry. The data and structural break tests provide historical insights into the nonstationary dynamics of these price variables, known as the container crisis. We find that the container crisis was a specific event in the shipping market. Our data shed light on the industry dynamics of shipping cartels. We leave a detailed analysis based on these data for future work. 

\textbf{Acknowledgement} \\
We benefited from anonymous referees and participants at the International Association of Maritime Economists 2022. We thank Akimitsu Ashida and Hiroyuki Sato for sharing industry knowledge and expertise as ex-chairperson in the 1980s. And we thank Mikio Tasaka, Yasuhiro Fujita, Jong-khil Han and Yutaka Yamomoto for professional comments. This study was supported by JSPS KAKENHI Grant Numbers 20K22129 and 22K13501.

\bibliographystyle{aer}
\bibliography{ship_bib}

\appendix
\begin{landscape}
{
\begin{table}[htb]\centering
{\tiny{}
  \begin{tabular}{lllll}
  Container freight rate & Year & Source & Measured Units & Note\\\hline
               & 1965,1970,1975,1979 & Issues of Our Ocean Shipping &dollars per 100 tons-mile&Liner sector, three main cargo\\
               & 1973-1976 & Current status of marine transportation &converted dollars per TEU&Revenue and quantity on only Transpacific and Asia-Europe routes\\
               &1976-1994 & Global Container Markets Drewry Shipping Consultants &dollars by FEU &Missing 1976-1989 in Asia-Europe\\
               &1994-2009 & Review of Maritime Transport &dollars by TEU&All market-level info is available\\
              & & & &\\
  Liner freight rate & & & &\\\hline
               &1965-2009 & Review of Maritime Transport &index rate based on 1995(=100)&Global liner freight rate\\
               & & & &\\
  Quantity & & & &\\\hline
                    &1966-1972 & Containerization International 1973&TEU& Aggregate carrying capacity for each route\\
                    &1970,1974,1978,1980 &The Container Crisis 1982 &1 million ton&
                    Aggregate quantity for each route\\
                    &1975,1978,1981,1984,1987,1990&Container transportation cost and profitability 1980/2000 &1000TEU&Aggregate quantity for each route\\
                    &1985-1997&World Sea Trade Service &1000TEU&All market-level info is available\\
                    &1994-2009 & Review of Maritime Transport &1000TEU& All market-level info is available\\
                     & & & &\\
                    &(1970,1975,1977,1978,1979) &*Issues of Our Ocean Shipping &D/W tons&Aggregate quantity for each route\\
                    &(1973)& *Containerization International 1975 &1000TEU&Aggregate quantity for each route\\
                    &(1978,1981)&*The Container market to 1990 &TEU&Aggregate carrying capacity for each route and container type\\
                    &(1983)&*World Container Data 1985 &1000TEU&Aggregate quantity for each route
  \end{tabular}
  \begin{tablenotes}
\item[a]\textit{Note:} The data sources with ($*$) are used for reference to check the consistency of the trend.
\end{tablenotes}
  }
  \caption{Overview of data sources.}
  \label{tb:overview_of_datasources}
\end{table}
\begin{figure}[!ht]
\begin{minipage}[b]{0.45\linewidth}
  \centering
  \includegraphics[height = 0.4\textheight]{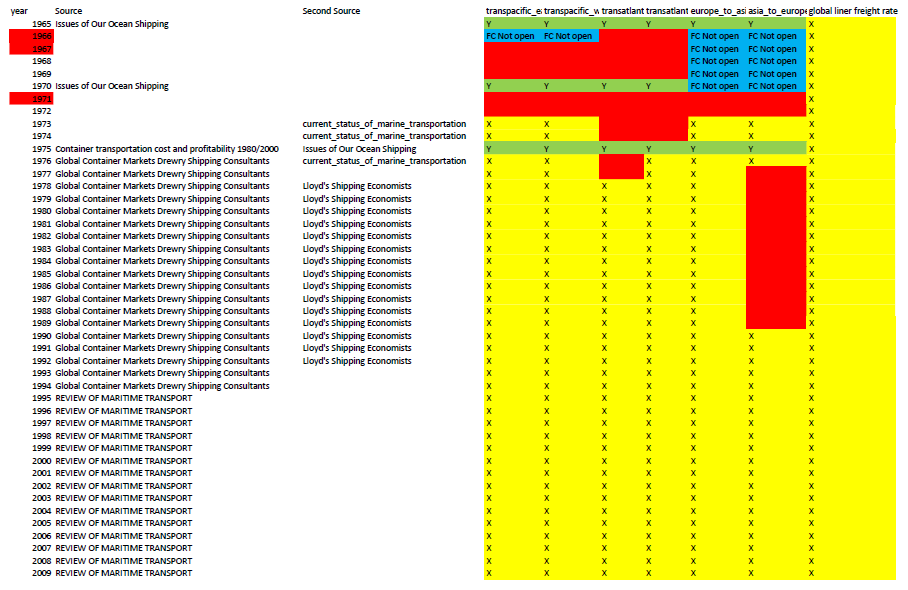}
  \end{minipage}
  \begin{minipage}[b]{0.45\linewidth}
  \centering
  \includegraphics[height = 0.4\textheight]{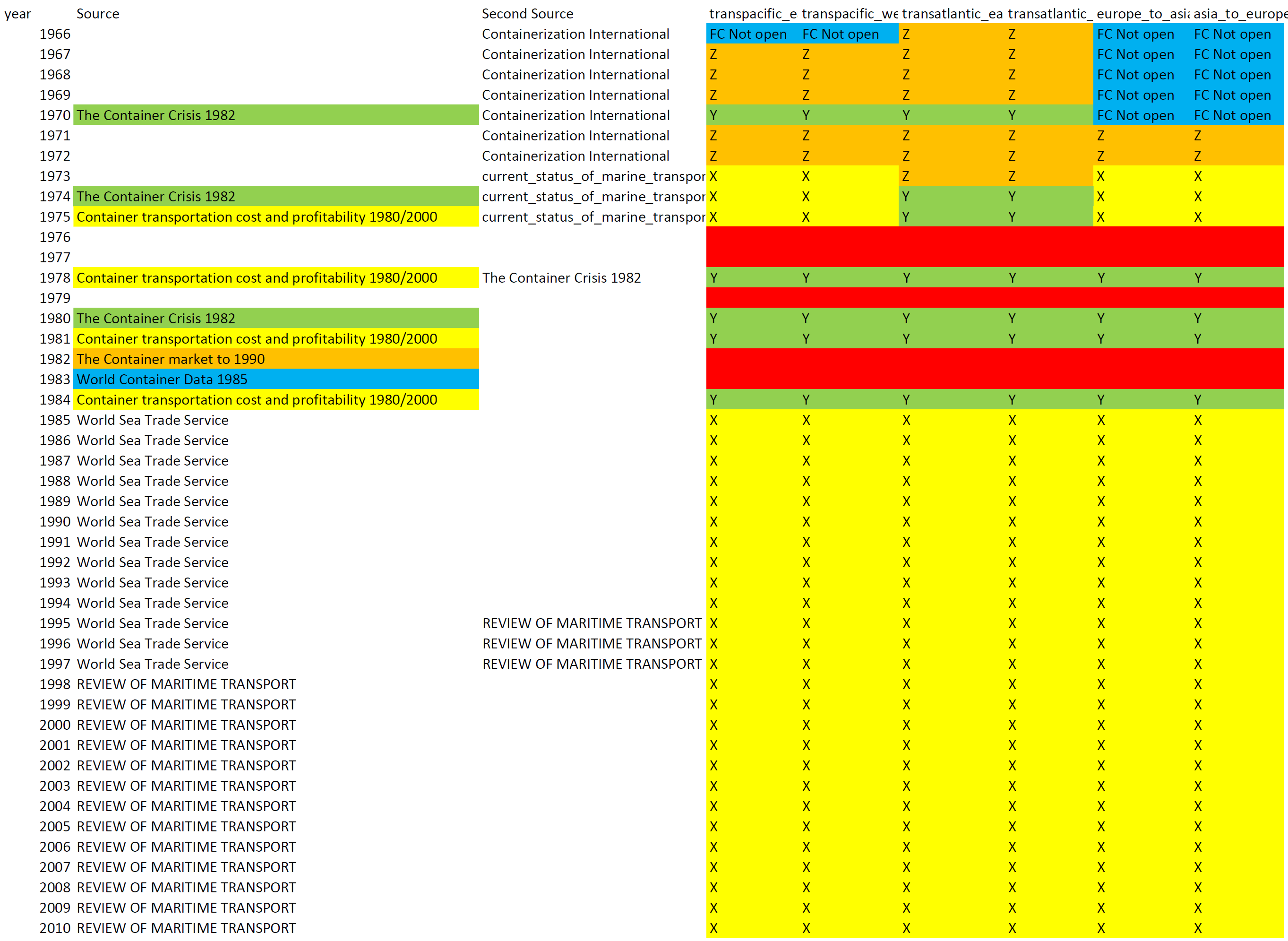}
  \end{minipage}
\caption{The observed and imputed variable cells for market-level data.}
{\tiny{}
\begin{tablenotes}
\item[a]\textit{Note:} In the yellow X cells, the route-level variables are recorded in data. In the green Y cells, the route-level variables that cannot identify the eastbound and westbound routes are recorded, so that we assign the values based on the fixed ratio of the eastbound to the westbound in the closest period. In the orange Z cells, although route-level variables are not recorded, ship-level variables with targeted routes are available, so that we convert firm-level information into route-level variables based on the fixed ratio in the closest period. The red cells are missing values. Thus, we impute the values via interpolation based on the observed points in each route and global liner freight rate.
\end{tablenotes}
}
\end{figure}
}
\end{landscape}

\section{Details of global containerization  (Not for publication)}\label{sec:details_of_global_containerization}

\paragraph{Transforming the cost structure.}
In the 1960s, the volume of cargo movement on ocean liners and port handling costs increased, putting pressure on the management of shipping companies. However, in the U.S., the cargo was loaded into standardized containers for transport, which not only eliminated the need for onboard cargo handling equipment but also enabled cargo handling in the rain. In addition, cargo handling in container shipping has saved a significant amount of work done by workers. Furthermore, they significantly reduce the time and cost required for cargo handling at ports. However, containerization has made the industry more capital-intensive.

\paragraph{Lowering barriers to entry.}
The development of containerization has dramatically lowered barriers to entry for new shipping companies. Prior to the development of containerization, large and small cargoes had to be loaded and unloaded individually, which required skilled stacking techniques and expertise to prevent cargo collapse during the voyage and a significant amount of time and workforce for cargo handling and management. As containerization has progressed, cargo sizes have been standardized, and many stacking techniques and expertise have become unnecessary, significantly reducing the time and labor required for cargo handling.

\paragraph{Rise of (non-conference) shipping companies in developing countries.}
Several national and local governments competed to build container terminals, which resulted in the loss of technological and equipment advantages of existing shipping companies. Thus, barriers to entry for new shipping companies that did not have the technology or bases for loading and unloading dropped significantly.\footnote{Further, containerization has made it possible to quickly reload cargo onto freight trains and trailers, which has led to the rapid development of international intermodal transportation. In the past, shipping companies were contracted to transport goods only by sea between ports and were not involved in inland transportation. The relationship between inland and vessel transportation is investigated by \cite{bernhofen2016estimating} and \cite{levinson2016box} explained it anecdotally. } In parallel with the development of containerization, developing and socialist countries began to develop the ocean shipping business as a core industry, and state-owned shipping companies in these countries began to enter the market as non-conference members.

\paragraph{Shifting from ``closed" conferences to ``open" conferences.}
Furthermore, conferences came to be strongly perceived as an impediment to the entry of the shipping industries of developing countries into the world trade market, mainly in the UNCTAD arena. In response to this trend, the United Nations Convention on a Code of Conduct for Linear Shipping Conferences was concluded in 1974. This convention allows shipping companies from trading parties to participate in shipping conferences. Moreover, when a conference determines the ratio of cargo transportation by shipping companies, the ratio of origin-destination shipping companies to third-country shipping companies should be 4:4:2. In the 1980s, conferences related to European routes began to ask non-conference companies to join conferences to prevent their market share from declining.\footnote{Theoretically, in the 1970s, the potential entrants in the conference markets consisted of non-containerized liner shipping companies in the shipping conferences. In the 1980s, the potential entrants added non-conference firms. However, the number of non-conference firms joining shipping conferences is marginal.}

\paragraph{Forming a consortium.}
As containerization progressed, container routes required considerable investment in ship construction and container terminal ownership. Therefore, multiple companies began to form consortiums to reduce the scale of investment while maintaining a certain level of service.\footnote{For example, these consortia include the TRIO Group, formed by European shipping lines, and the Ace Group, consisting of Japanese, Korean, French, and other shipping lines.} Specifically, consortiums are engaged in business co-operation such as space chartering, joint use of container terminals, and coordination of operation schedules.\footnote{ Major container shipping companies are forming global alliances such as THE Alliance and the Ocean Alliance. They are utilizing a consortium framework.} Both ``conference" and ``consortium" are defined as concerted actions by competition laws in various countries, which then require exemptions from competition laws. However, member companies at a conference agree on freight rates, and consortium members do not have any consent. Individual shipping companies in the consortium determine freight rates and sales activities. Thus, the consortium is not treated as a single firm.

\section{Data construction (Not for publication)}\label{sec:data_construction}

The container shipping industry is a fascinating laboratory for investigating industry dynamics. Because the industry started global shipping in 1966, we can determine the initial state of the market's dynamic structure. There is also substantial firm entry and exit in the industry, which makes it ideal for studying industry dynamics in the global markets. Finally, the markets observe the dynamics of shipping alliances, mergers, and consolidations. 

Despite its significance, a panel dataset regarding the container freight rate and shipping quantity on the three major trade routes (front-haul and back-haul separately) between 1966 and 2009 was not available.\footnote{Similar to related papers focusing on the container freight rate panel data, \cite{luo2009econometric} use shipping demand and freight rate data between 1980 and 2007. As for shipping demand, they used the world container throughput reported in the Drewry Annual Container Market Review and Forecast. The container freight rate is calculated as the weighted average of Transpacific, Europe-Far East and Transatlantic trades from the same data source. Because of the data limitation, they calculated the missing period (1980–1993) from the General Freight Index in the Shipping Statistics Yearbook 2007, using a simple statistical equation between the container freight rate and the general freight index from 1994 to 2008.} Although, unified instructions and guidance to construct the dataset are provided by published books and available data, the construction and merging processes involve conversion and imputation errors from multiple datasets. Table \ref{tb:overview_of_datasources} presents the data sources. Instead of manually constructing the freight index from the commodity-level freight rate via formal but complicated processes, we adopted at tractable imputation approach, which links multiple data sources that overlap information for a few years. The corresponding code is also provided to ensure that an interested researcher can replicate the construction. 

Collecting data on container freight rates and shipping quantity, particularly before 1994, is not trivial because there is no single data source. In the subsequent subsections, we provide detailed data construction for each container freight rate and shipping quantity.

\subsection{Container freight rate}

To construct the data regarding the container freight rate, we refer to the data sources presented in Table \ref{tb:overview_of_datasources} and the liner freight rate before global containerization.

\subsubsection{Shipping conference and Liner freight rate}\label{subsec:shipping_conferences}

The liner shipping industry has traditionally formed shipping conferences including explicit cartels. Until 1984, shipping conferences played an important role in the container-shipping market. For example, in February 1967, a container shipping firm, Matson, started an operation on the Atlantic route under the freight rate contracted with one of conferences, the TPFC. Thus, the container freight rate was determined by liner shipping conferences.

\begin{figure}[!ht]
\begin{center}
\includegraphics[height = 0.4\textheight]{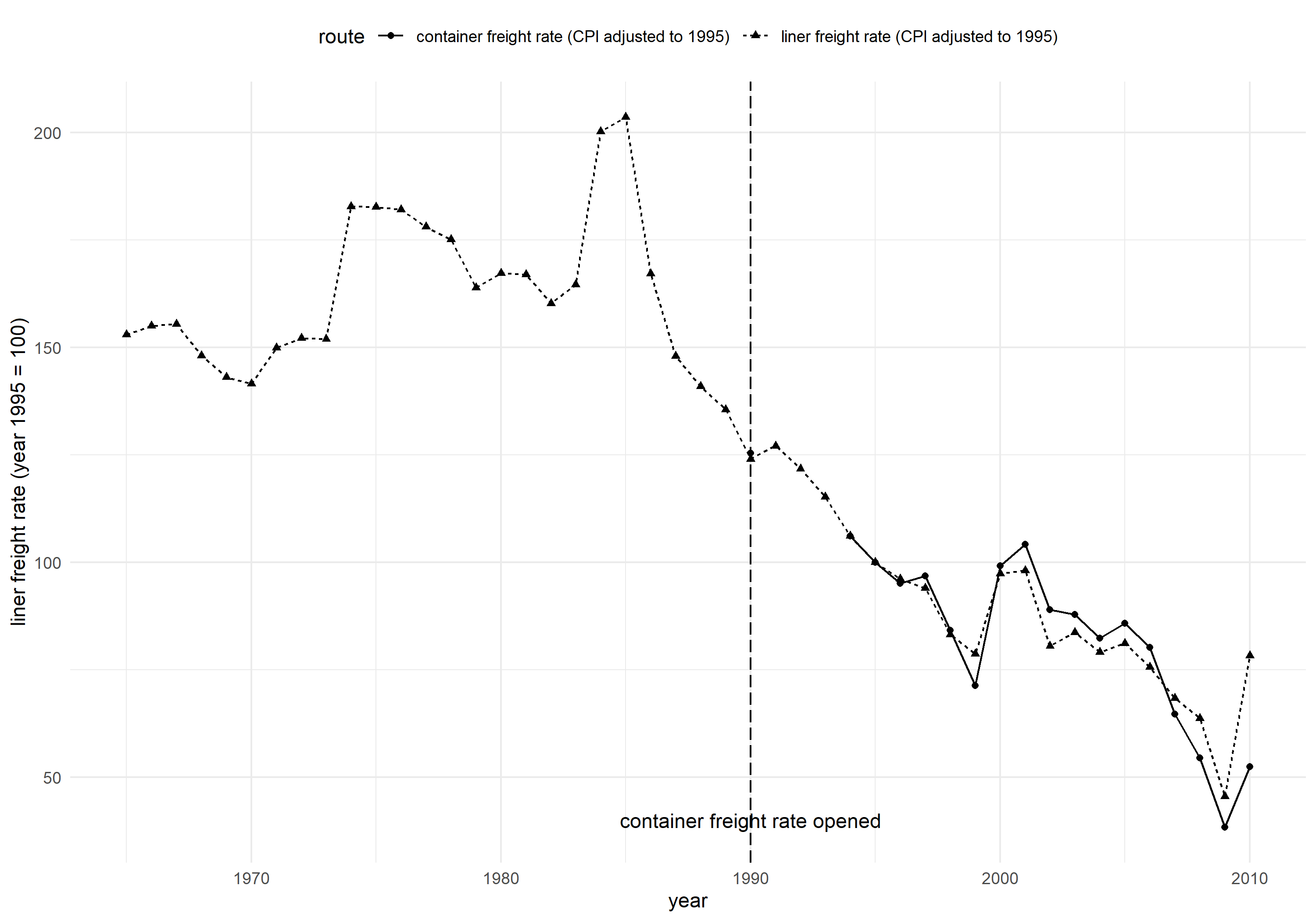}
\end{center}
\caption{The trend of the liner freight rate.}
\begin{tablenotes}
\item[a]\textit{Note:} Source: \textit{Review of Maritime Transport}, published annually by UNCTAD. The measure of the price index is not explicitly mentioned in the Review of Maritime Transport.
\end{tablenotes}
\label{fg:liner_freight_rate}
\end{figure}

This unique feature allows us to use the liner freight rate to approximate the container freight rate. Figure \ref{fg:liner_freight_rate} compares the liner freight rate with the container freight rate recorded since 1995. In particular, “Issues of Our Ocean Shipping” recorded that in 1978, 76.0\% of U.S. liner transports, 66.4\% of European liner transports, and 68.8\% of Asian and Australian liner transports were containerized. Therefore, more than two-thirds of the liner freight rate was determined by the container freight rate in the 1970s and the 1990s.

\subsubsection{Container freight rate between 1966 and 1975}

Data on freight rates between 1966 (i.e., the beginning of the industry) and 1975 were missing from a single data source. Thus, we need to infer and impute data from multiple sources and information in subsequent years. Fortunately, we can use the institutional knowledge in Section \ref{subsec:shipping_conferences}, the liner freight rate, and the published data that overlap the data in subsequent years in multiple data sources.

The ``Issues of Our Ocean Shipping" (\textit{``Wagakuni no Gaikou Kaiun Ni Tsuite", in Japanese}) records the freight rate of liner shipping on three major routes in 1965, 1970, 1975, and 1979. The freight rates for the three types of cargo (electric appliances, clothes, and ceramic products) are available. The freight rates were measured in dollars for 100 tons per mile. Shipping miles between specific ports are listed.\footnote{For the Transpacific and Asia-Europe routes, the shipping miles between specific ports, Japan-Hamburg and Japan-San Francisco routes are mentioned. However, only for the transatlantic route, the specific ports were not mentioned. Thus, we take the average of the shipping miles for Hamburg-San Francisco (i.e., west coast) and Hamburg-Halifax routes (i.e., east coast).} Using this information, we can recover route-level container freight rates for 1965, 1970, and 1975. Finally, we assumed that the proportions of liner and container freight rates for each route were fixed before 1975. Under this assumption, we recovered and interpolated the eastbound and westbound container freights for each route based on the fixed proportion of the liner and container freight rates and that of subsequent years, which overlaps container freight rates in multiple data sources.\footnote{The overlapped information in 1979 is key for merging multiple data before and after 1975. However, the freight rate swung irregularly and non-proportionally in both data sources. For making the smooth transition of the freight rate in the merged dataset, we calculated the conversion rate based on the freight rate of 1975 from the \textit{Issues of our ocean shipping} and that of 1976 from \textit{Global container markets year}.}

\subsubsection{Container freight rate between 1976 and 1994}

The data on the freight rate between 1976 and 1994 were recorded in the \textit{Global Container Markets Drewry Shipping Consultants}. The recording of the freight rate started in 1976 for the transatlantic and Transpacific routes and in 1990 for the Europe-Asia route. Thus, data on the freight rate of the Europe-Asia route between 1976 and 1989 are missing.\footnote{\textit{Global Container Markets Drewry Shipping Consultants} stated that, ``the Europe-Far East trade is the least well documented of the axial routes (page 109)".} 

For the imputation of the missing data, we assumed that the proportion of liner and container freight rates for eastbound and westbound Asia-Europe routes were fixed between 1976 and 1990. This assumption implies that there is no time-varying difference between the eastbound and westbound Asia-Europe routes. Under this assumption, we recovered the container freight rates of the eastbound and westbound Asia-Europe routes.

\subsubsection{Container freight rate after 1995}\label{subsec:freight_rate_after_1995}

The freight rate data were recorded in the \textit{Review of Maritime Transport} which refers to the \textit{Containerization International Yearbook}. The recording of freight rates began in June 1994.\footnote{\cite{jeon2022learning} stated: ``The first dataset on firm-level investment and capital is therefore supplemented with the historical price and quantity data compiled from the Review of Maritime Transport published by the United Nations that goes back to 1997." (p. 9). However, we confirm that the data in 1994 are available in \textit{Containerization International Yearbook}.} The \textit{Containerization International Yearbook} mentioned that \textit{``the information is derived mainly from confidential reports provided by some of the main carriers and other public sources."} \textit{Review of Maritime Transport} is often used because of the consistency of data sources and public availability. For example, \cite{jeon2022learning} used quarterly freight data.

\begin{figure}[!ht]
\begin{center}
\includegraphics[height = 0.5\textheight]{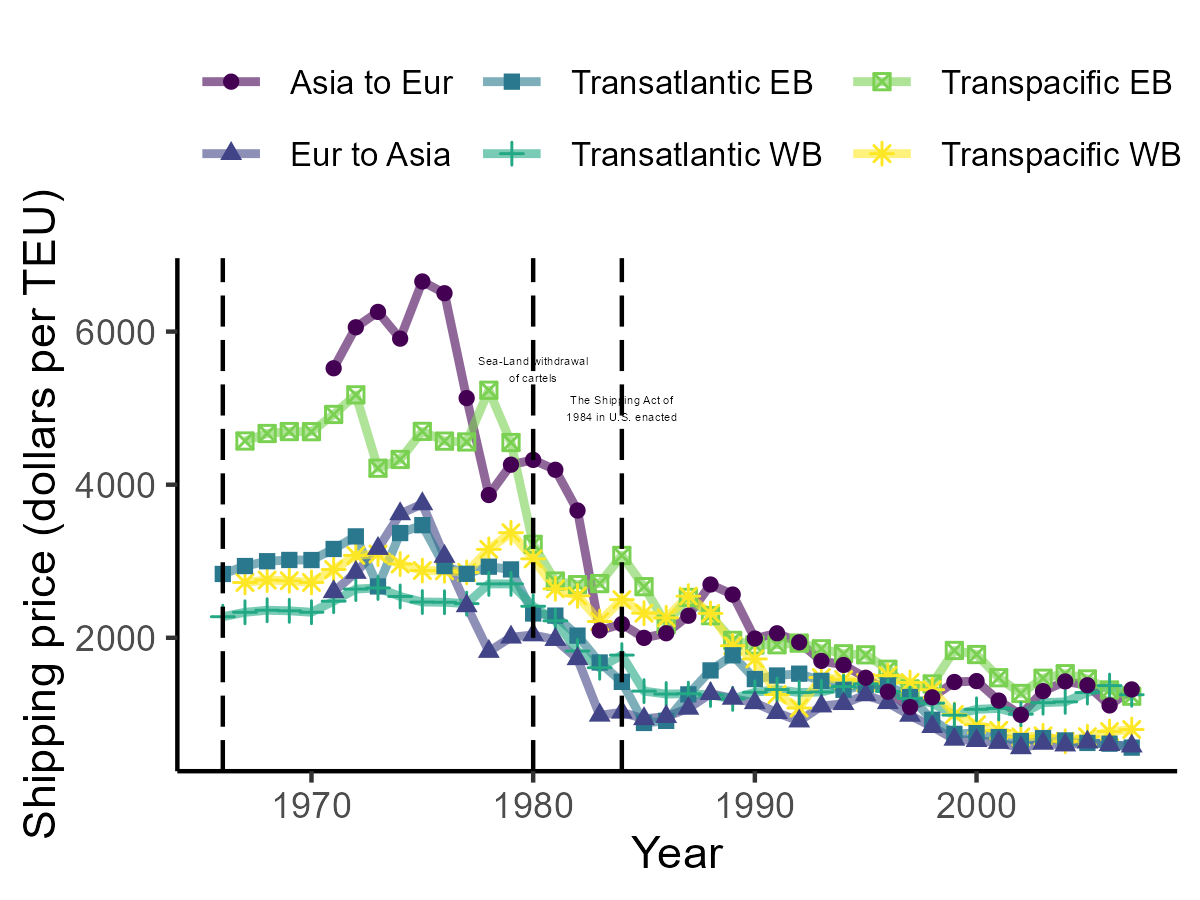}
\end{center}
\caption{The trend of the container freight rate.}
\label{fg:container_freight_rate_each_route}
\end{figure}

The above steps generate time-series data of the container freight rate for each route between 1966 and 2009. Figure \ref{fg:container_freight_rate_each_route} shows the trend of the container freight rate (CPI-adjusted to 1995) with the liner freight rate depicted in Figure \ref{fg:liner_freight_rate}. In particular, the recovered and imputed data points in the 1960s and the 1970s quantitatively capture anecdotal and institutional evidence. First, for the transatlantic and Transpacific routes, a peak in 1976 and a sharp declining trend in 1979 were observed. Second, the container freight rate on each route corresponds to fluctuations in the liner freight rate.

\subsection{Shipping quantity}

To construct data on container trade volume, we refer to the five data sources shown in Table \ref{tb:overview_of_datasources}.

\subsubsection{Shipping quantity between 1966 and 1975}

Official data on shipping quantities between 1966 and 1969 were not available. However, we could keep track of the development of the container vessels during this period. First, data regarding vessel capacity were recorded in \textit{Containerization International 1973}.\footnote{``Theory and Practice of Container Shipping" (\textit{``Container Yusou no Riron to Jissai", in Japanese}) contains information about the launch of container ships owned by the United States' four main firms, Moore-McCormack, U.S. Lines, SeaLand, and CML (American Export Isbrandtsen). This helps us to identify the exact launch year of full-container ships by the largest players, that is, United States' container shipping companies. We use the data source only to check institutional evidence.} This information identifies the carrier composition of the Transatlantic, Transpacific, and Asia-Europe routes between 1966 and 1972.

Second, the \textit{Review of Maritime Transport} (1971) summarizes the relationship between annual global container carrying capacity and container capacity in 1969 and 1970 on the transatlantic route. We assume that container capacities were fully exploited during this period. This assumption is not restrictive, because the demand for container shipping was considerably high during this period. Under this assumption, the shipping quantity is recovered by calculating the total shipping quantity from the container capacity and invariant conversion rate.


Third, the data on shipping quantity in 1970 and 1974 were recorded in \textit{The Container Crisis 1982}, and the data for 1973 were recorded in \text{Containerization International 1975}. However, data for 1971 and 1972 were missing. The data contains regional levels of container shipping, that is, container traffic in Asia, North America, and Europe. One alternative is the imputation approach, which employs external data that provide information about missing data.\footnote{\cite{jeon2022learning} stated that: “We set the start date for firms’’ information as the second quarter of 1966, which is the date of the first international container voyage. Then, we employed quarterly data on the value of trade by origin-destination pair from the IMF Direction of Trade Statistics database to impute the missing data on demand states from 1966: Q2-1996: Q4.” (p. 25), and “To translate the value of trade to the quantity of container trade, the demand state for the 1997–-2014 period was regressed on the de-trended value of trade. Then, the demand states for periods with missing data are constructed as predicted values from the regression. For the 1997–-2014 period, actual demand states are used.” (footnote 37).} We assign the recorded trade volume to each route in proportion based on subsequent years.\footnote{Alternatively, we could assign the aggregate container shipping quantity to westbound and eastbound routes based on the data on the value of trade by origin-destination pair from \textit{the IMF Direction of Trade Statistics database}. We did not take the full-imputation approach because the value of trade includes goods transported by the bulk shipping service.} For missing years, we interpolated the trade volume using the observed data points.

\subsubsection{Shipping quantity between 1976 and 1994}

Data on shipping quantity between 1976 and 1994 were recorded in \textit{World Sea Trade Service}, \textit{Container transportation cost and profitability 1980/2000}, \textit{The Container Crisis 1982}, and \textit{World Container Data 1985}. \textit{World Sea Trade Service} started recording data in 1985, whereas the other data sources recorded the data before 1985 but irregularly. Specifically, the quantity data for 1971, 1972, 1973, 1976, 1977, and 1979 were missing for all routes, and the quantity data for 1970, 1974, 1980, 1982, and 1983 were recorded in the total shipping quantity, summing the eastbound and westbound for each market.

\subsubsection{Shipping quantity after 1995}
Data on shipping quantity after 1995 were recorded in the \textit{Review of Maritime Transport} and \textit{World's sea trades}. The former refers to \textit{World Sea Trade Service Review, various issues, 1996}; \textit{Journal of Commerce, various issues, 1996}; and \textit{Containerization International, various issues, 1996.} As in Section \ref{subsec:freight_rate_after_1995}, shipping quantity data are easily constructed from a single data source, the \textit{Review of Maritime Transport}. Before merging the data before and after 1995, we confirm that the discrepancy between the two data sources is marginal for almost all market-year observations. However, only the quantities on the transatlantic eastbound and westbound routes in 1995 were somewhat different.

\begin{figure}[!ht]
\begin{center}
\includegraphics[height = 0.5\textheight]{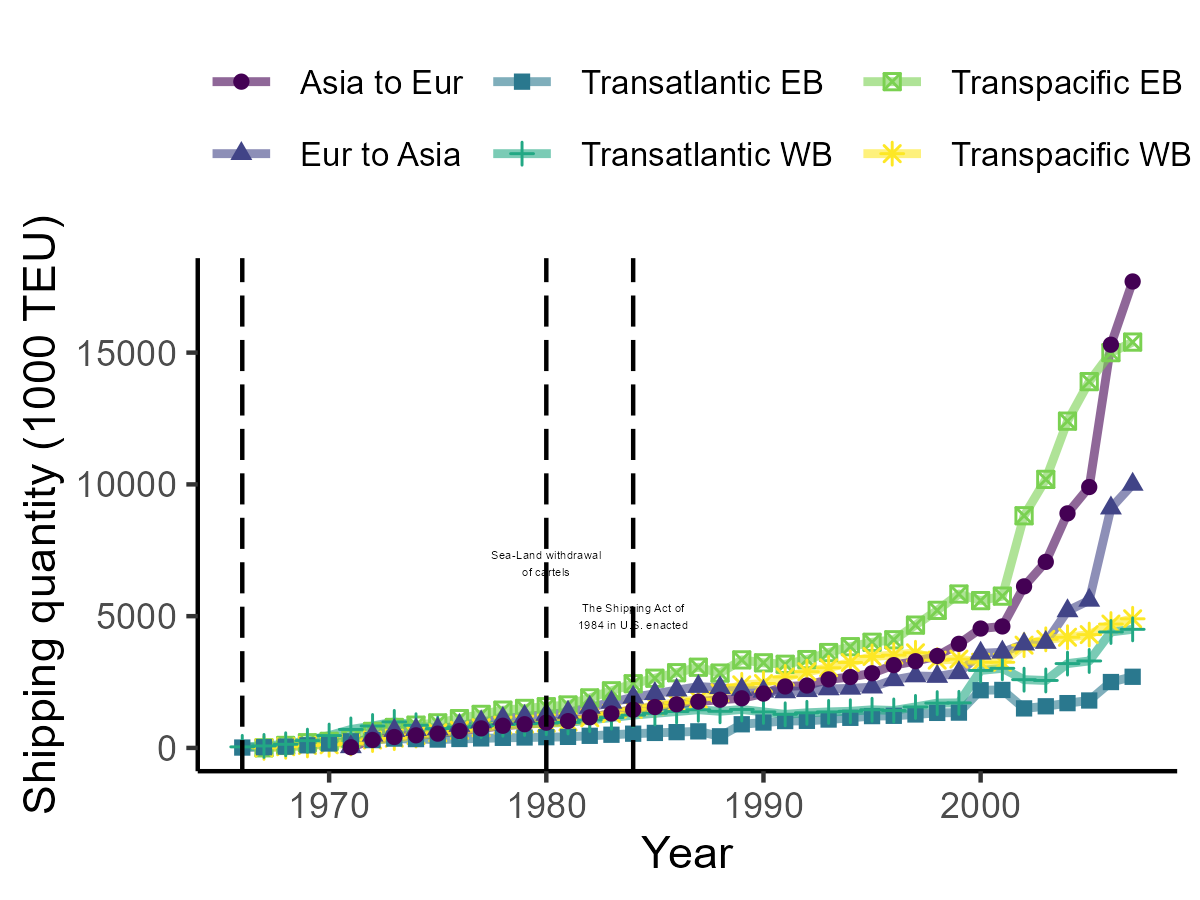}

\end{center}
\caption{The trend of the container shipping quantity. The trend before 1976 is based on ship-level carrying capacity information on each route.}
\label{fg:container_shipping_quantity_each_route}
\end{figure}

The above steps generate time-series data of container shipping quantity measured by 1000 TEU for each route between 1966 and 2009. Figure \ref{fg:container_shipping_quantity_each_route} illustrates the trend in the container shipping quantity. Shipping quantity on all routes increased monotonically between 1973 and 2000. After 2000, this trend increased exponentially, particularly on the Transpacific eastbound and Europe-to-Asia routes.

\subsection{Newbuilding, Secondhand, and Scrap prices}
We provide recommendations for constructing newbuilding, secondhand, and scrap prices for container ships from 1967. Our data were collected from a series of \textit{Review (1971-1998)} published by Fearnley and \textit{Lloyd's Shipping Economist (1983-1990)} published by Lloyd's of London Press. 

\subsubsection{Newbuilding prices}
We collected industry-year-level newbuilding prices per 18000 DWT bulk ships from \textit{Review (1971-1998)}). Using the overlapped year, we converted prices into consistent prices under the assumption that the conversion rate is invariant across years. Then, we divided the price per 12000 dwt by 10 to convert it to the price per 1200 TEU and then converted it into the price per TEU. Finally, we convert bulk prices into container prices based on \textit{Lloyd's Shipping Economist (1983-1990)}.

\subsubsection{Secondhand prices}
We collected industry-year-level secondhand prices per 16000 DWT CSD (liner type) ships from \textit{Review (1971-1998)}. Using the overlapped year, we converted prices into consistent prices under the assumption that the conversion rate is invariant across years.

We assume a fixed depreciation rate ($= X$) and a fixed conversion rate ($= a$) for 1981 and 1983 and observe the following:
\begin{align*}
    1981: 2.7 + 2X &= 16.0a \quad\text{(18-year depreciation)}\\
    1983: 0.8 + 0X &=11a \quad\text{(20-year depreciation)}
\end{align*}
We solved the equations for $X$ and $a$. To obtain the price per $1600dwt+15X$, we multiplied the above price by 12000/16000 and obtained the price per 12,000dwt. We then divided the price by 10 to convert it to the price per 1200TEU. We then converted it into the price per TEU of 5 years ship by dividing it by 1200. Finally, we converted liner prices into container prices based on \textit{Lloyd's Shipping Economist (1983-1990)}.

\subsubsection{Scrap prices}

We collected industry-year-level scrap prices per LTD in the Far-East. Using the overlapped year, we converted prices into consistent prices under the assumption that the conversion rate is invariant across years. We then divided each LTD by four to obtain the price per dwt, multiplied it by 10, and converted it into the price per TEU.\footnote{The rate is based on \textit{https://nippon.zaidan.info/seikabutsu/2002/00264/contents/030.htm}.}

\end{document}